
%


\input harvmac.tex


%
\def\ubrackfill#1{$\mathsurround=0pt
	\kern2.5pt\vrule depth#1\leaders\hrule\hfill\vrule depth#1\kern2.5pt$}
\def\contract#1{\mathop{\vbox{\ialign{##\crcr\noalign{\kern3pt}
	\ubrackfill{3pt}\crcr\noalign{\kern3pt\nointerlineskip}
	$\hfil\displaystyle{#1}\hfil$\crcr}}}\limits
}

\def\ubrack#1{$\mathsurround=0pt
	\vrule depth#1\leaders\hrule\hfill\vrule depth#1$}
\def\dbrack#1{$\mathsurround=0pt
	\vrule height#1\leaders\hrule\hfill\vrule height#1$}
\def\ucontract#1#2{\mathop{\vbox{\ialign{##\crcr\noalign{\kern 4pt}
	\ubrack{#2}\crcr\noalign{\kern 4pt\nointerlineskip}
	$\hskip #1\relax$\crcr}}}\limits}
\def\dcontract#1#2{\mathop{\vbox{\ialign{##\crcr
	$\hskip #1\relax$\crcr\noalign{\kern0pt}
	\dbrack{#2}\crcr\noalign{\kern0pt\nointerlineskip}
	}}}\limits
}

\def\ucont#1#2#3{^{\kern-#3\ucontract{#1}{#2}\kern #3\kern-#1}}
\def\dcont#1#2#3{_{\kern-#3\dcontract{#1}{#2}\kern #3\kern-#1}}



\font\tenmsy=msbm10
\font\sevenmsy=msbm10 at 7pt
\font\fivemsy=msbm10 at 5pt
\newfam\msyfam 
\textfont\msyfam=\tenmsy
\scriptfont\msyfam=\sevenmsy
\scriptscriptfont\msyfam=\fivemsy
\def\blackB{\fam\msyfam\tenmsy}
\def\Z{{\blackB Z}}

\def\M{{\cal M}}

\let\d\partial

\let\s\sigma
\let\L\Langel
\let\R\rangle

\def\frac#1#2{{\textstyle{#1\over #2}}}

\def\eqalignD#1{
\vcenter{\openup1\jot\halign{
\hfil$\displaystyle{##}$~&
$\displaystyle{##}$\hfil~&
$\displaystyle{##}$\hfil\cr
#1}}
}
\def\eqalignT#1{
\vcenter{\openup1\jot\halign{
\hfil$\displaystyle{##}$~&
$\displaystyle{##}$\hfil~&
$\displaystyle{##}$\hfil~&
$\displaystyle{##}$\hfil\cr
#1}}
}

\def\eqalignS#1{
\vcenter{\openup1\jot\halign{
\hfil$\displaystyle{##}$~&
$\displaystyle{##}$\hfil~&
$\displaystyle{##}$\hfil~&
$\displaystyle{##}$\hfil~&
$\displaystyle{##}$\hfil~&
$\displaystyle{##}$\hfil~&
$\displaystyle{##}$\hfil\cr
#1}}
}

\def\text#1{\quad\hbox{#1}\quad}

\def\la{\lambda}
\def\e{\epsilon}

\def\y{{\infty}}
\def\vp{{\varphi}}

\let\Rw\Rightarrow

\def\rw{\rightarrow}

\def\Om{\mathop{\Omega}\limits}

\def\L{\langle}
\def\R{\rangle}

\def\lra{\leftrightarrow}
\def\su{\widehat{su}}
\def\osp{\widehat{osp}}

\def\M{{\cal M}}

\def\S{{\cal S}}

\newcount\eqnum
\eqnum=0
\def\eq{\eqno(\secsym\the\meqno)\global\advance\meqno by1}
\def\eqlabel#1{{\xdef#1{\secsym\the\meqno}}\eq }

\newwrite\refs
\def\startreferences{
 \immediate\openout\refs=references
 \immediate\write\refs{\baselineskip=14pt \parindent=16pt \parskip=2pt}
}
\startreferences

\refno=0
\def\aref#1{\global\advance\refno by1
 \immediate\write\refs{\noexpand\item{\the\refno.}#1\hfil\par}}
\def\ref#1{\aref{#1}\the\refno}
\def\refname#1{\xdef#1{\the\refno}}

\newcount\exno
\exno=0
\def\Ex{\global\advance\exno by1{\noindent\sl Example \the\exno:

\nobreak\par\nobreak}}

\parskip=6pt

\overfullrule=0mm

\def\frac#1#2{{#1 \over #2}}

\let\d=\partial

\def\suh{{\widehat {su}}}
\def\uh{{\widehat u}}
\def\rw{{\rightarrow}}

\def\tC{{\widetilde C}}

\def\tB{{\widetilde B}}

\newwrite\refs
\def\startreferences{
 \immediate\openout\refs=references
 \immediate\write\refs{\baselineskip=14pt \parindent=16pt \parskip=2pt}
}
\startreferences

\refno=0
\def\aref#1{\global\advance\refno by1
 \immediate\write\refs{\noexpand\item{\the\refno.}#1\hfil\par}}
\def\ref#1{\aref{#1}\the\refno}
\def\refname#1{\xdef#1{\the\refno}}


 \hbox{DCPT-05/29}

\Title{\vbox{\baselineskip12pt
\hbox{ }}}
{\vbox {\centerline{A quasi-particle description  of the $\M(3,p)$ models}}}

\smallskip
\centerline{ P. Jacob and P. Mathieu
}
\smallskip\centerline{ Department of Mathematical Sciences, University of Durham, Durham, DH1 3LE, UK}
\centerline{and}
\centerline{D\'epartement de
physique,Universit\'e Laval, Qu\'ebec, Canada G1K 7P4}
\smallskip\centerline{(patrick.jacob@durham.ac.uk, pmathieu@phy.ulaval.ca)}
\vskip .2in
\bigskip
\bigskip
\centerline{\bf Abstract}
\bigskip
\noindent The $\M(3,p)$ minimal models are reconsidered from the point of view of the extended algebra whose generators are the energy-momentum  tensor and  the primary field $\phi_{2,1}$ of dimension $(p-2)/4$.
Within this framework, we provide a quasi-particle description of these models, in which all states are expressed solely in terms of the $\phi_{2,1}$-modes. More precisely, 
we show that all the states can be written in terms of $\phi_{2,1}$-type highest-weight states and their $\phi_{2,1}$-descendants. We further demonstrate that the conformal dimension of these highest-weight states can be calculated from the $\phi_{2,1}$  commutation relations, the highest-weight conditions and associativity. For the simplest models $(p=5,\,7)$, the full spectrum is explicitly reconstructed along these lines. For $p$ odd, the commutation relations between the $\phi_{2,1}$ modes take the form of infinite sums, i.e., of  generalized commutation relations akin to parafermionic models. In that case, an unexpected  operator, generalizing the Witten index, is unravelled in the OPE of $\phi_{2,1}$ with itself.
A quasi-particle basis formulated in terms of the sole $\phi_{1,2}$ modes is  studied for all allowed values of $p$. We argue that it is governed by jagged-type partitions
further subject a difference 2 condition at distance 2. 
We demonstrate the correctness of this basis by constructing its generating function, from which the proper fermionic expression of the combination of the Virasoro irreducible characters $\chi_{1,s}$ and $\chi_{1,p-s}$ (for $1\leq s\leq [p/3]+1$) are recovered.
As an aside, a practical technique for implementing associativity at the level of mode computations is presented, together with a general discussion of the relation between associativity and the Jacobi identities.


\Date{05/05\ }


\newsec{Introduction}

\subsec{Quasi-particle description of  the minimal models: extended algebra  vs spinon-type formulation}

Fermionic-type character expressions are known for all Virasoro minimal models.\foot{A brief review of the origin of fermionic characters in conformal field theory, together with
an extended list of  references (focusing mainly on the early works), is presented in the introduction of [\ref{P. Jacob and
P. Mathieu,  Nucl. Phys. {\bf B 620} (2002)
351.}\refname\JMb].
Note that the first fermionic  character formulas appeared in mathematics [\ref{J. Lepowsky and
M. Primc,  Contemporary Mathematics {\bf 46} AMS, Providence, 1985.}] and its
 interpretation in terms of
an exclusion principle seems to go back to [\ref{J. Lepowsky and R.L. Wilson, Proc. Nat. Acad. Sci. USA {\bf 78} (1981) 7254.}].
Another early reference on fermionic formulas is [\ref{B.L. Feigin, T. Nakanishi  and H. Ooguri,   Int. J. Mod. Phys.
{\bf A7} Suppl. {\bf 1A} (1992) 217.
}\refname\FNO].  This subject has somewhat exploded with [\ref{R. Kedem, T.R. Klassen, B. M. McCoy and E. Melzer, Phys. Lett. {\bf B304} (1993) 263 and  Phys. Lett. {\bf B307} (1993) 68; R. Kedem, B. M. McCoy and E. Melzer, in {\it Recent progress in statistical mechanics and quantum field theory},
ed. by P. Bouwknegt et al, World Scientific (1995) 195. }\refname\KKMM]  where various
fermionic-sum formulas were proposed and interpreted in terms of a generalized exclusion principle. More
formulas pertaining to the Virasoro minimal models were conjectured and proved in [\ref{E. Melzer, Int. J. Mod. Phys. {\bf A9} (1994) 1115; A. Berkovich, Nucl. Phys. {\bf B241} (1984) 333.}] and [\ref{A. Berkovich and B.M. McCoy,  Lett. Math. Phys. {\bf 37} (1996) 49; A. Berkovich,  B.M. McCoy and A. Schilling, Comm. Math Phys.
{\bf 191} (1998) 329.}\refname\Ber]. The works of the
last reference rely heavily on a special truncation of the spin chain
XXZ spectrum. A different method for proving the Virasoro fermionic formulas has been developed in [\ref{S.O. Warnaar, 
 J. Stat. Phys. {\bf 82} (1996) 657 and J. Stat. Phys. {\bf 84} (1997) 49.}], using the
connection with the RSOS models [\ref{G.E. Andrews, R.J. Baxter and P.J. Forrester, J. Stat. Phys. {\bf 35} (1984) 193.}]. The recent work  [\ref{T. Welsh, {\it Fermionic expressions for minimal model Virasoro characters}, to appear in
Memoirs of the American Mathematical Society, math.CO/0212154.}\refname\Wel]  contains exhaustive results and many
references to other works. A
different approach to the construction of fermionic formulas  has been initiated in [\ref{B. Feigin and A.
V. Stoyanovsky, {\it Quasi-particles models for the representations of Lie algebras and geometry of 
flag manifold}, hep-th/9308079.}]
and vigorously extended recently by Feigin and collaborators.} 
However, most of these Virasoro characters, as well as  the underlying basis of states, have not been explained within a conformal field theoretical set up. To look for such an intrinsic understanding of the fermionic characters is a fundamental quest. So far, only the minimal models $\M(2,p)$ have been successfully addressed from that perspective [\FNO, \ref{B.L. Feigin and E. Frenkel, Adv. Sov. Math. {\bf 16} (1993) 139.}\refname\FF]. But is there any natural lines of attack for handling this question?
Two potential avenues can readily been identified: a reinterpretation  in terms of  an extended algebra or  a spinon-type reformulation.

All minimal models have a hidden extended conformal symmetry. The extension is obtained by adding one generator to
the usual energy-momentum tensor. This extra generator is
$\phi_{1,p-1}= \phi_{p'-1,1}$,\foot{The field $\phi_{p'-1,1}$ has dimension $h_{p'-1,1}=(p-2)(p'-2)/4$, which   is integer when either  $p$ or $p'$ is of the form $ 2+4m$. In that case, this extended algebra is at the roots of the corresponding A-D type block-diagonal modular invariants [\ref{A. Capelli, C. Itzykson and J.-B. Zuber, Comm. Math. Phys {\bf 13} (1987) 1.}]  (for this interpretation,  see for instance [\ref{G. Moore and N. Seiberg, Nucl. Phys {\bf B313} (1989) 16.}]). Notice that even for this case, the minimal models have not been reformulated  in terms of the  representation theory of this extended algebra. But we stress that such a reformulation does not requires $h_{p'-1,1}$ to be integer (a point that is plainly illustrated in this work).} with fusion rule
$$\phi_{p'-1,1}\times \phi_{p'-1,1}= \phi_{1,1}=I\; . \eqlabel\exta$$
Reformulating the minimal models from the perspective of this  two-generator extended algebra has the obvious 
disadvantage  that the defining extended  algebra differs from one  model to the other. But on the other hand, that the algebraic formalism is tailor-made for each model strongly suggests the existence of an underlying basis, generated by the $\phi_{p'-1,1}$ modes, that would be free of singular vectors, i.e., a quasi-particle basis.


An alternative formulation might be considered, 
in which the fundamental spectrum generating field is either  $\phi_{1,2}$ or $\phi_{2,1}$. 
In such a formulation, the extended algebra (incorporating one of $\phi_{1,2}$ and $\phi_{2,1}$)  would generically be
`non-abelian' (that is, multi-channels), e.g., since 
$$\phi_{2,1}\times \phi_{2,1} = \phi_{1,1} + \phi_{3,1}\;,\eqlabel\fus$$
with the two fields on the right-hand side having dimensions that do not differ by an integer.\foot{They differ by
an integer when $p'=2$ but in that case $\phi_{3,1}$ is a descendant of $\phi_{1,1}$ [\ref{P. di Francesco and P. Mathieu, Phys. Lett. {\bf B270} (1992) 79.}].} 
In that framework, the added  algebra generators would be the set of fields $\{\phi_{r,1}\}$ in which $\phi_{2,1}$ plays the role of a basic generator. This is much like the parafermionic algebra [\ref{A.B. Zamolodchikov and V.A. Fateev, Sov. Phys. JETP {\bf 43} (1985) 215.}\refname\ZF]  which is spanned by the parafermionic fields $\psi_n$, with $0\leq n \leq k-1$; the $\psi_n$'s can  all be obtained from the multiple  product of $\psi_1$ with itself. However, the parafermionic algebra is single-channel (that is, $\psi_n\times \psi_m= \psi_{n+m}$), in contradistinction with the  above $\{\phi_{r,1}\}$ algebra.
This yet-to-be-defined $\phi_{2,1}$-type reformulation of the minimal models  is actually closer to the spinon description of the $\su(2)_k$ WZW models  (see [\ref{D. Bernard, V.
Pasquier and D. Serban , Nucl. Phys. {\bf B428 } (1994) 612.},\ref{P. Bouwknegt, A. W.W. Ludwig and K. Schoutens,
Phys. Lett. {\bf B338} (1994) 448.}\refname\BPS] for $k=1$ and  [\ref{P. Bouwknegt, A. W.W. Ludwig and K. Schoutens,
 Phys. Lett. {\bf B359} (1995) 304}\refname\BLS] for $k>1$). In the WZW context, the spinon field refers to the primary doublet of spin $j=1/2$, denoted  $\phi_{1/2}$ and the analog of the above fusion rule is $\phi_{1/2}\times \phi_{1/2} = \phi_{0}+ \phi_{1}$ ($\phi_1$ being absent for $k=1$). Because of this analogy, this approach will  be referred to as a spinon-type reformulation.

 Extended algebras associated to a set of OPE including multi-channel ones of the 
sort (\fus) appear rather difficult to analyze, however. But that the field
$\phi_{1,2}$ could be regarded as a primary field in a formulation based on an extended algebra containing
$\phi_{2,1}$ (or the inverse) would fit nicely the following suggestive formula that would necessarily results
from such a construction
$${2 h_{1,2} h_{2,1}\over c_{p',p}} = {1\over 8}\; . \eqlabel\belle$$
This expression is  valid for all minimal models $\M(p',p)$, with
$$h_{r,s} = {(rp-sp')^2-(p-p')^2\over 4 pp'}\qquad{\rm and }\qquad  c_{p',p}= 1-{6(p-p')^2\over pp'}\;, \eq$$
($1\leq r\leq p'-1$ and $ 1\leq s\leq p-1$, with $p'<p$).  


Whether this spinon-type approach  can be worked out in general remains to be seen. In this work, we consider the special
case for which  this second method reduces to the first one,  namely the $\M(3,p)$ minimal models. When $p'=3$, the fusion of
$\phi_{2,1}$ with itself yields the identity:
$$\phi_{2,1}\times \phi_{2,1}  = I \;. \eqlabel\fus$$

 \subsec{Generalities on the structure of the $\phi$-algebra}

Our aim here is thus to  describe the $\M(3,p)$ models in terms of the extended
algebra associated to the OPEs
$$
\eqalign{
\phi(z)\phi(w)=& {1\over (z-w)^{2h}} \left[ I+(z-w)^2 {2h\over c} T(w)+\cdots \right]\S \;, \cr
T(z) \phi(w) = & {h\phi(w)\over (z-w)^2}+{\d\phi(w)\over (z-w)}+\cdots \cr
T(z) T(w) = & {c_{3,p}/2\over (z-w)^4}+ {2T(w)\over (z-w)^2}+{\d T(w)\over (z-w)}+\cdots \cr}
\eqlabel\opep$$
with $$\phi\equiv \phi_{2,1}\qquad {\rm and} \qquad h\equiv h_{1,2}= {p-2\over 4}\;,\eq$$
 instead of through the
 irreducible representations 
 of the Virasoro algebra. In the first OPE, we have included an operator $\S$ which is enforced (see below) by the mutual locality of $\phi$ for  the cases where $h\not\in \Z_+/2$, that is,  for $p$ odd; $\S$ anticommutes with $\phi$ and commutes with $T$.
 That  this
algebra is associative [\ref{A.L. Zamolodchikov, Theo. Math. Phys. {\bf 63} (1985) 1205.}\refname\Zam], at least for the value central charge  $c_{3,p}$, is guaranteed by the associativity of the operator algebra of 
minimal models [\ref{A.A.
Belavin, A.M. Polyakov and A.B. Zamolodchikov, Nucl. Phys. {\bf B241} (1984) {333}.}\refname\BPZ, \ref{Vl.S. Dotsenko and V.A. Fateev,
Nucl. Phys. {\bf B235} (1984) 312;  Nucl. Phys. {\bf
B251} (1985) 691;  Phys. Lett. {\bf 154B} (1985)
291.}\refname\DFc].

Note that, as for the  WZW models [\ref{V.G. Knizhnik and A.B. Zamolodchikov, Nucl. Phys.  {\bf B247} (1984) 83.}\refname\KZ] or the parafermionic models [\ZF], the Virasoro algebra lives in the `enveloping algebra' of $\phi$.  This is well-known for  the free-fermionic formulation of the Ising model, which  corresponds to the $p=4$  case of the above construction (i.e., $\psi=\phi_{1,3}=\phi_{2,1}$).  That $T$ might not occur in a (pole-type) singular  term  in the OPE $\phi(z)\phi(w)$  (which is the case for $p<6$) is no obstruction to the fact that $T$ can always be written as a bilinear in the field $\phi$.  In that sense, the `$\phi$-algebra' is essentially defined by the first OPE in (\opep) and it will be understood as such.

We stress  that, in writing (\opep), we do not consider the generic form of the OPE $\phi(z)\phi(w)$ [\Zam].  In fact, we have imposed an implicit restriction on those terms that can appear on the right-hand side of (\opep): these are solely the descendants of the Virasoro  identity. In particular, no bilinear composite of $\phi$ does occur. This condition is restrictive; it actually fixes $c$ to  finitely many values (and in particular, $c$ cannot be free). For instance, when $\phi$ has dimension $h=3/2$, there are only two solutions to the associativity conditions: $c=7/10 $ and $c=-21/4$.

The structure of the $\phi$-algebra (\opep) depends crucially upon the parity of $p$.
For $p$ even, $2h$ is  integer so that $\S=I$. There are then only integer powers of $z-w$ on the right-hand side of (\opep).  This results into ordinary (anti)commutation relations. The basic data for the first few models with $p$ even are: 
$$\eqalignS{
& \M(3,8): \; &h=3/2,\; &c=-21/4\phantom{0}: \;\simeq {\cal SM}(2,8)\;,\qquad & \M(3,16): \quad &h=7/2,\quad &c=-161/8 \;,\cr
& \M(3,10): \; &h=2,\; &c=-44/5\phantom{0}: \;\simeq  [{\cal M}(2,5)]^2 \;,\qquad & \M(3,20): \quad &h=9/2,\quad &c=-279/10 \;, \cr
& \M(3,14): \; &h=3,\; &c=-114/7: \;\simeq  W_3(3,7) \;, \qquad  &\M(3,22): \quad &h=5,\quad &c=-350/11 \;.\cr
}\eq$$
 ${\cal SM}$ stands for a superconformal minimal model. That associativity could fix the central charge in spite of the fact that these relations might satisfy the  Jacobi identities for generic values of the central charge, as in the first three cases [\Zam],  is a point that is clarified in the appendix.
The solution displayed in the last three cases  is among those few (3 for $h=7/2$ and 5 for the other two cases) found in
an investigation of low-dimensional two-generators extended
algebras [\ref{P. Bouwknegt, Phys. Lett. {\bf B207} (1988) 295.}]. 

On the other hand, for $p$ odd,  $h=\pm 1/4$ mod 1. That case presents at once a severe complication in that
the commutation relations resulting from (\opep) take the form of infinite sums, as for parafermionic models
[\ZF]  (cf. section 2).  Our analysis of
the first two models in this class, $h=3/4$ and $h=5/4$, shows that in these cases $c$ is uniquely determined and it agrees with $c_{3,5}$ and $c_{3,7}$  respectively.  But that two solutions are found for the next
case ($h=9/4$) suggests that this uniqueness is not generic.

As previously mentioned, the analysis  of the $p$ odd case also reveals that  the first OPE in (\opep) requires the introduction  of the operator $\S$, anticommuting with $\phi$. This operator can be viewed as a generalization of the Witten operator $(-1)^F$ which anticommutes with
fermions [\ref{E. Witten, Nucl. Phys {\bf B202} (1992) 253.}]. Here we have $\S= (-1)^{p {\cal F}} $ with 
$$\phi\times \phi = (-1)^{p {\cal F}} \, I\qquad {\rm with} \qquad  (-1)^{p {\cal F}}\, \phi = (-1)^p \phi \, (-1)^{p {\cal F}}\;. \eqlabel\SF$$
This is quite similar to the phase factor introduced in the $\suh(2)_1$  commutation relations for the spinon fields (for which  $h=1/4$) in [\BLS] (cf. eq. (3) there). The  existence of this operator is definitely forced by associativity. Note that it does not appear in the OPE of the physical field $\Phi(z,{\bar z})= \phi(z)\phi({\bar z})$ since it squares to 1.  Since $T$ is bilinear in $\phi$, it commutes with $\S$.

\subsec{The spinon-type reformulation of the $\M(3,p)$ models: setting the problem}

What does a $\phi$-algebra reformulation of the  $\M(3,p)$ models should amount to? It should certainly allow us to fix completely the spectrum of the model  for a given $p$, that is, to determine the highest-weight states and their conformal dimension.  The highest-weight states turn out to be completely characterized by an integer $\ell$ such that $0\leq \ell\leq (p-2)/ 2$. The highest-weight state conditions are formulated directly in terms of the $\phi$-modes and they read
$$\phi_{-h-n+\frac\ell2}\, |\s_\ell\R= 0 \qquad n>0\;. \eqlabel\hwc$$
We then observe that the conformal dimension of some highest-weight states $|\s_\ell\R$ are directly obtained from the commutation relations deduced from (\opep) and the highest-weight conditions (\hwc). However, it is generally necessary to invoke associativity to fix the full spectrum. This analysis is performed in section 2 
for  the two simplest $\M(3,p)$ models apart from the Ising case, namely $\M(3,5)$ and $ \M(3,7)$.
The obtained conformal dimensions confirm the following identification with the Virasoro highest-weight states: $|\phi_{1,s}\R= |\s_{s-1}\R$ for $ 1\leq s<p/2$.


 A second required ingredient is a complete characterization of  the descendant states in terms of the action of the $\phi$-modes.  Note in that regard that a single $\phi$-module  generically decompose into the direct sum of two Virasoso modules. Each Virasoro component will be singled out  from  the subset of descendant states containing an even or odd number of $\phi$-modes acting on the highest-weight state. This feature is well known  for those
$\M(3,p)$ models that have previously been formulated from that perspective, namely  the $\M(3,4),\,  \M(3,5) $ and
$\M(3,8) $ models, for which
$\phi$ is respectively the free fermion,  the fundamental graded parafermion [\ref{P.
Jacob and P. Mathieu, Nucl. Phys. {\bf B630} (2002) 433}\refname\JM]
 and the superpartner of the energy-momentum tensor. 
In the present case, the Virasoro highest-weight states 
$|\phi_{1,s}\R= |\s_{s-1}\R$ 
and  $|\phi_{1,p-s}\R$ ($1\leq s<p/2)$  are
 combined into a single
$\phi$-module since $|\phi_{1,p-s}\R$ is  a descendant of $|\phi_{1,s}\R$:
$$|\phi_{1,p-s}\R = \phi_{-h+(s-1)/2}|\phi_{1,s}\R \;. \eqlabel\desce$$
This matches  the relation
$$h_{1,p-s}-h_{1,s}= h-{(s-1) \over 2}\;.\eq$$

We stress that we invoke here only very general aspects of a yet-to-be-defined representation theory of the $\phi$-algebra. Actually, we make use the notion of a highest-weight state,  that of $\phi$-lowering operators and associativity.

 Ideally, we would also like to look  for a complete description of the irreducible modules, i.e., a basis of states. There are two natural possibilities for such a basis, one formulated in terms of the combination of the Virasoro modes and the $\phi$-modes or one formulated solely in terms of the $\phi$-modes. 
In this work, we focus on the second possibility.  In that case,  the basis is expected to be of a quasi-particle type.
 A quasi-particle basis entails the  construction of the Hilbert space by the action of the quasi-particle creation
operators  subject to a restriction rule, i.e., a filling process controlled by an exclusion principle.
 The obtention of this basis is our main result.

\subsec{Quasi-particle basis of states}

In the quest for a quasi-particle basis, we were guided by our previous construction [\JM] of the quasi-particle basis for the graded
parafermions [\ref{J. M. Camino, A. V. Ramallo and
J. M. Sanchez de Santos, Nucl.Phys. {\bf B530} (1998) 715.}\refname\CRS] in terms of jagged partitions.  Such partitions differ from
standard partitions, for which parts are non-increasing from left to right, in that  a possible increase between  parts at short
distance is allowed.  

Manipulations with the generalized commutation
relations derived from (\opep) together with  simple considerations on the structure of the highest-weight modules
lead us to infer  the general form of a candidate basis (section 3).  It takes the following form. The highest-weight  modules $ |\s_\ell\R$ are described by the successive action of the lowering $\phi$-modes subject to specific constraints. In the $N$-particle sector, with strings of lowering modes written in the form
$$ \phi_{-h+\frac\ell2+\frac{(N-1)}2-n_1} \, 
\phi_{-h+\frac\ell2+\frac{(N-2)}2-n_{2}} \cdots
 \phi_{-h+\frac\ell2+\frac12-n_{N-1}}\,
  \phi_{-h+\frac\ell2-n_N}\, |\s_\ell\R\;,
 \eqlabel\struT$$ 
these constraints are:
$$\eqalignD{
 & n_i\geq n_{i+1}-r+1\qquad 
  & n_i\geq n_{i+2}+2\cr
&  n_{N-1} \geq \ell-r\qquad  & n_N\geq 0 \;,\cr}\eqlabel\jagG$$
where
$$2r= p-5\;.\eq$$   
The $n_i$'s are integers for $p$ odd and alternate between integer and half-integer values when $p$ is even (from right to left). The condition  $0\leq \ell\leq k$ is linked to the boundary condition on $n_{N-1} $ that appears to be `complete' (i.e., to represent the full set of  conditions that singles out the different modules)  only in these cases.

This quasi-particle basis of states was known in at least three cases: $p=4,5$ and $8$. In each case, it reduces to (\jagG). For $p=4$, the quasi-particle basis is that of a free fermion:
$$b_{-s_1}\cdots b_{-s_N}|0\R\qquad {\rm with }\qquad s_i\geq s_{i+1}+1 \; . \eq$$
 To rephrase this in terms of our previous notation, we set
$$ s_i = n_i+\frac12 -{(N-i)\over 2}\;\qquad \Rw\qquad n_i \geq n_{i+1}+\frac32\;, \eq$$ which is indeed the first condition in (\jagG) when $r=-1/2$. In this special case, the second condition is a consequence of the first one. 

For $p=5$, as mentioned previously, the model is the simplest example of a graded $\Z_k$ parafermion [\JM], corresponding to the value $k=1$. The quasi-particle basis in that special case, when reformulated in terms of the $n_i$'s reads (see [\JM], end of section 5):
$$n_i\geq n_{i+1}+1\;. \eq$$ This is again the first condition in (\jagG) for $r=0$; here again it  implies the second condition of  (\jagG).

For $p=8$, it has been pointed out that $\M(3,8) \simeq {\cal SM}(2,8)$. Now, we have recently obtained the quasi-particle basis of superconformal models ${\cal  SM}(2,4\kappa)$ in [\ref{J.-F. Fortin, P. Jacob and P.
Mathieu, J. Phys. A: Mat. Gen. {\bf 38} (2005) 1699.}\refname\FJMa]. It is expressed solely in terms of $G$ modes ($G$ being the superpartner of $T$) and for $\kappa=4$, it takes the form
$$G_{-s_1}\cdots G_{-s_N}|0\R\qquad {\rm with }\qquad s_i\geq s_{i+1}-1\;  \qquad {\rm and}\qquad  s_i\geq s_{i+2}+1 \; , \eq$$ with all $s_i$ half-integers (cf. [\FJMa] eqs (13) and (17)).
 The relation between $s_i$ and $n_i$ being
$s_i = n_i +3/2-{(N-i)/2}$, the above conditions translate into
$$n_i\geq n_{i+1}-\frac12 \qquad {\rm and }\qquad n_i\geq n_{i+2}+2\;. \eq$$
We again recover (\jagG) for $r=3/2$. 


\subsec{Fermionic characters}

The complete module of $|\s_\ell\R$ is obtained by summing over all these states (\struT) satisfying (\jagG) and  all values of $N$. Granting the correctness of this  basis, one can then enumerate states in highest-weight
modules (section 4). This leads to 
a fermionic expression of the  (normalized) character that takes the simple form
$$\hat{\chi}_\ell(q)= \sum_{m_1,m_2,\cdots m_k\geq 0} {q^{{\bf m}\, {B} \, {\bf m} + C\cdot {\bf m}}
\over (q)_{m_1}\cdots (q)_{m_k}}\;, \eqlabel\genfa$$
where the matrices $B$ and $C$ are  defined in section 4. In terms of the Virasoro characters, $\hat{\chi}_\ell$ decomposes as follows.:
$$
\hat{\chi}_{\ell}(q)= q^{-h_{1,s}+c/24} \left[\chi^{{\rm Vir}}_{1,s}(q)+ q^{h_{1,p-s}-h_{1,s}}\chi_{1,p-s}^{{\rm
Vir}}(q)\right]\; \quad  ( \ell=s-1\; 0\leq\ell\ [p/3] )\;.\eq$$
We recover in this way the fermionic sums given in  [\ref{A.G. Bytsko, J. Phys. A: Math. Gen. {\bf 32} (1999)
8045; A.G. Bytsko and A.
Fring, Comm. Math. Phys. {\bf 209} (2000) 179.}\refname\By], where these expressions where first conjectured. The expression of some of
these characters can also be found in [\Ber,
\ref{G. Andrews, Pacific J. Math. {\bf 114} (1984) 267.},
\ref{O. Foda and Y.-H.Quano,  Int. J. Mod. Phys. A12 (1997) 1651.}\refname\FQ] (see also the second reference of [\KKMM]). Their derivation from the general expressions in [\Wel] is presented in  [\ref{B. Feigin, O. Foda and T. Welsh, {\it Andrews-Gordon identities from combinations of Virasoro characters}, math-ph/0504014.}\refname\FFW]. This equivalence confirms the correctness of the basis  (\struT)-(\jagG).

Obtaining the  fermionic character amounts to  finding the generating function for all the states (\struT)-(\jagG).  But finding such generating functions is in general a difficult problem. In the present  case, we modify the characterization of our states in order to make use of a related generating function derived in [\ref{B. Feigin, M. Jimbo and T. Miwa, {\it Vertex operator algebra arising from the minimal series
 $M(3,p)$ and monomial basis}, in MathPhys odyssey, Prog. Math. Phys., 23,
    Birkh\"auser (2002) 179.}\refname\FJM]. Given that we read off our generating function (up to boundary terms) from [\FJM] and that this latter article
explicitly deals with an algebra related to the $\M(3,p)$ models,  it is appropriate to clarify the
relation between this work and the present one. The authors of [\FJM] construct a vertex operator algebra
out of the product of the $\M(3,p)$ model and a free boson. The algebra is generated by the two local fields:
$a(z) = V_1 \phi_{2,1}$ and $a^*(z)=V_{-1} \phi_{2,1}$, where $V_{1}$ and $V_{-1}$ are vertex operators with 
dimension such that 
$a,\, a^*$ have respective dimension $1$ and $p-3$.  The monomial basis underlying this vertex operator algebra is
spanned by the modes $a_\la = (a_{\la_1},\cdots  ,a_{\la_N})$, with the $\la_i$'s subject to $\la_i\geq \la_{i+2}+2r$.  The origin of
the exclusion here is rooted in polynomial relations of the type $a^2\d  ^{n}a=0$ for $0\leq n\leq p-3$ (plus an
additional one).  In view of our results, by stripping off the contribution of the free boson, we end up with the
jagged-type basis (\jagG). In preparing the revised version of this work, we became aware of [\ref{B. Feigin, M. Jimbo, T. Miwa, E. Mukhin and Y. Takayama,  {\it Set of rigged paths with Virasoro characters}, math.QA/0506150.}\refname\FJMMT] where the translation of these results to the $\M(3,p)$ models is  performed  and the resulting basis (cf. Lemma 5.5 there) agrees perfectly with ours for $\ell=0$.

\newsec{The $\M(3,p)$ algebra for $p$ odd }

\subsec{Generalized commutation relations}

Let us first justify the necessity of the operator $\S$ for $p$ odd  by invoking associativity.
Consider a correlator of the form $\L \phi(z_1)\phi(z_2)\phi(z_3)\cdots \R$. The first  OPE in (\opep) shows that moving $\phi(z_2)$ and then $\phi(z_3)$ in front of $\phi(z_1)$ induces a negative phase (since $2h$ is half-integer for $p$ odd):
$$\L \phi(z_1)\phi(z_2)\phi(z_3)\cdots \R = -\L \phi(z_2)\phi(z_3)\phi(z_1)\cdots \R \sim - {1\over z_{23}^{2h} } \, \L \left(\S\ +\cdots \right) \phi(z_1)\cdots \R \,,
\eq$$
which is to be compared with 
$$\L \phi(z_1)\phi(z_2)\phi(z_3)\cdots \R \sim  {1\over z_{23}^{2h}}\,  \L \phi(z_1) ( \S
\ +\cdots ) \cdots \R \;. \eq$$
The compatibility of  these expressions forces
$$ \S
 \, \phi(z) = -\phi(z)\, \S
 \;,\eq$$
which captures  the whole
effect of $\S
$ and allows for the identification with $(-1)^{p{\cal F}}$ previously displayed in (\SF).
For  the $\M(3,5)$ case, this operator also
appears in the description of the model as a graded parafermionic theory. This is briefly reviewed in the following subsection.

Consider now the commutation relations associated to (\opep). 
The mode decomposition of $\phi$ acting on a state of `charge' $\ell$ (or, in the sector specified by
the integer $\ell$) is:
$$\phi(z)= \sum_{n\in\Z} z^{-n-\frac\ell2}\; \phi_{-h+\frac\ell2+n} \;. \eq$$
The field $\phi$ itself has charge 1.  A first useful commutation relation is obtained  by evaluating the
following integral:
$${1\over (2\pi i)^2}\oint dw \oint dz \, \phi(z)\, \phi(w)\, z^{\frac\ell2+n}\, w^{\frac\ell2+m-1}\, (z-w)^{2h-1} \; ,\eqlabel\iut$$
in two different ways [\ZF]. 
That yields
$$\sum_{t=0}^\y C^{(t)}_{2h-1} [ \phi_{\frac\ell2+n-t+h}\,  \phi_{\frac\ell2+m+t-h}+
\phi_{\frac\ell2+m-1-t+h} \, \phi_{\frac\ell2+n+1+t-h}] = \S
 \, \delta_{n+m+\ell,0} \;, \eqlabel\genca$$ 
where 
$$ C_u^{(t)}=  {\Gamma(t-u)\over t!\,
\Gamma(-u)} \;.\eq$$
If we replace 
$$w^{\frac\ell2+m-1}(z-w)^{2h-1}\quad \rw\quad w^{\frac\ell2+m+1}(z-w)^{2h-3} \eq$$
in (\iut), in order to pick up
the contribution of $T$ (the change in the power of $w$ being purely conventional),  we get instead
$$\eqalign{
\sum_{t=0}^\y C^{(t)}_{2h-3} &[ \phi_{\frac\ell2+n-2-t+h}\,  \phi_{\frac\ell2+m+t-h+2}+
\phi_{\frac\ell2+m-1-t+h}\, \phi_{\frac\ell2+n+1+t-h}]\cr &  =
\frac12\left(\frac\ell2+n\right)\left(\frac\ell2+n-1\right)\S
\, 
\delta_{n+m+\ell,0} +
\frac{2h}{c}L_{n+m+\ell}\, \S
\;. \cr} \eqlabel\gencb$$ 
In the following, we will refer to the above  two generalized commutation relations as follows:
$$\eqalign{
& (\genca) \; \lra \; {\rm I}_{n,m,\ell}\cr
& (\gencb) \; \lra \; {\rm II}_{n,m,\ell} \;. \cr}\eqlabel\superno$$

Let us first test the relative sign for the two terms in the infinite sum (\gencb)\foot{Note that this sign is
correlated  to that in (\genca). This computation verifies the `bosonic' nature of the field $\phi$ within the
present framework, i.e., that the interchange of the two fields in (\iut) does not generate a minus sign.} by acting with both sides of (\gencb) on  the vacuum state
$|0\R$ (so that
$\ell=0$) which is such that
$$\phi_{-h+n}\, |0\R= 0 \qquad n>0 \;.\eqlabel\vacuhw$$
With $n=2$ and $ m=-2$, only one term contributes from the first sum and none from the second sum, so that
$$\phi_{h}\phi_{-h} |0\R= \S
|0\R \;.\eq$$
Taking instead $n=m=0$, only one term of the second sum contributes and we get the same result, confirming thus the
positive relative sign.

 
We have just stated in (\vacuhw) the highest-weight condition pertaining to the vacuum state. Let us denote by
$|\s_\ell\R$ the highest-weight state in the sector labeled by $\ell$.
Its highest-weight state characterization is
$$\phi_{-h-n+\frac\ell2}\, |\s_\ell\R= 0 \qquad n>0\;. \eqlabel\hwc$$
Note that in order for the dimension of the first descendant $\phi_{-h+\frac\ell2} |\s_\ell\R$ to be non-negative,
we require
$$0\leq \ell\leq {p-2\over 2} \;.\eqlabel\bobo$$
This bound will be assumed to hold from now on. The action of $\S
$ on a  highest-weight state  $|\s_\ell\R$ is normalized as
$$ \S
 |\s_\ell\R= |\s_\ell\R  \;.\eqlabel\zerov$$

A somewhat remarkable feature of the $\phi$-algebra for $p$ odd is that the dimension of the highest-weight state
$|\s_1\R$ follows directly from (\gencb) and (\hwc), exactly as for all parafermionic highest-weight states [\ZF]. (This, of course, is also true for $\s_0=I$.)
Applying both sides of (\gencb)  on
$|\s_1\R$  with
$n=m=0$, we see that no term contributes on the left-hand side, so that using (\zerov) and 
$L_0|\s_1\R= h_1 |\s_1\R$ we obtain
$${2h h_1\over c}= {1\over 8} \;, \eqlabel\belun$$ which is a special case of (\belle) when $c=c_{3,p}$ (and recall
that
$h=h_{2,1}$).  In other words, with  $c=c_{3,p}$, this relation fixes the value of $h_1$ to $h_{1,2}$.

As previously pointed out,  within the framework of this reformulation of the $\M(3,p)$ models in terms of the $\phi$-algebra,  a highest-weight $\phi$-module is generically a combination of
two Virasoro highest-weight modules (and this holds true irrespectively of the parity of $p$). Indeed, take for instance the vacuum state
$ |0\R=|\s_0\R= |\phi_{1,1}\R$; its first descendant will be $\phi_{-h}\, |0\R$, which is 
itself the Virasoro highest-weight state
$|\phi_{1,p-1}\R= |\phi_{2,1}\R$.  More generally, the states
$|\phi_{1,s}\R$ 
and  $|\phi_{1,p-s}\R$ will
 be combined into a single
module since $|\phi_{1,p-s}\R$ is  a descendant of $|\phi_{1,s}\R$ - cf. (\desce). Note also that 
$$\S
 |\phi_{1,p-s}\R = (-1)^p |\phi_{1,p-s} \R \;,\eq$$
still with the understanding that $s=\ell+1< p/2$.
 The only case for which the $\phi$-module reduces to a single Virasoro module is when $p$ is even and $s=p/2$.

\subsec{The $\M(3,5)$ model }

Let us first demonstrate, in a very explicit way, the necessity of the operator  $\S
$ anticommuting with $\phi$. For this, we evaluate
$\phi_{\frac14}\phi_{-\frac14}\phi_{-\frac34}|0\R$ in two different ways, symbolically written as:\foot{This is an  example of a mode-formulated associativity computation  -- cf.  the appendix for more detail.} 
$$\underbrace{\phi_{\frac14}\phi_{-\frac14}}_{{\rm I}_{-1,0,1}}\phi_{-\frac34}|0\R = 
\phi_{\frac14} \underbrace{\phi_{-\frac14}\phi_{-\frac34} }_{{\rm I}_{-1,0,0}}|0\R \;,\eq$$
using the notation introduced in (\superno).  This relation means that we commute the first two terms using
(\genca) with $n=-1,\, m=0,\, \ell=1$ ($\ell=1$ because we act on a state with $\ell=1$, namely
$\phi_{-\frac34}|0\R$) and we compare this with the result of commuting the second and third terms using again
(\genca) but now with $n=-1,\, m=0,\, \ell=0$. That leads to
$$\phi_{-\frac34}\phi_{\frac34}\phi_{-\frac34}|0\R + \S
 \phi_{-\frac34}|0\R = 0\;,\eqlabel\test$$
(i.e., $\phi_{-\frac14}\phi_{-\frac34}|0\R=0$ follows directly form ${\rm I}_{-1,0,0}$ and this is in agreement with
the absence of a level-one descendant in the vacuum module: $L_{-1}|0\R=0$). 
Next we use ${\rm
I}_{0,0,0}$ to obtain
$$\phi_{\frac34}\phi_{-\frac34}|0\R= \S
\, |0\R \;.\eq$$
The relation  (\test) becomes then 
$$\left[\phi_{-\frac34} \S
 +  \S
 \phi_{-\frac34} \right] |0\R = 0 \;,\eqlabel\testa$$
which is precisely what we wanted to establish: $\S
$ anticommutes with the modes of $\phi$. On the other hand, 
there is no way of fixing the eigenvalue of
$\S
$ on
$|0\R$ and it is thus chosen to be 1.


The central charge is obtained by evaluating 
$$\underbrace{\phi_{\frac14}\phi_{-\frac14}}_{{\rm II}_{-1,0,1}}\phi_{-\frac34}|0\R=
\phi_{\frac14} \underbrace{\phi_{-\frac14}\phi_{-\frac34} }_{{\rm II}_{-1,0,0}}|0\R \;.\eq$$
This leads to
$$-{3\over 8c} (5 c+3)\phi_{-\frac34}|0\R = 0 \;,\eq$$ fixing $c=-3/5$ as expected. Associativity is further tested by
computing the central charge as 
$$\underbrace{\phi_{-\frac34}\phi_{\frac34}}_{{\rm II}_{0,-1,1}}\phi_{-\frac34}|0\R=
\phi_{-\frac34} \underbrace{\phi_{\frac34}\phi_{-\frac34} }_{{\rm II}_{0,0,0}}|0\R \;.\eq$$
This gives the equivalent result:
$$-{1\over 8c} (7 c+9)\phi_{-\frac34}|0\R = \phi_{-\frac34}|0\R \;. \eq$$

Consider now the spectrum of the model. Since $p=5$, the bound (\bobo) yields $0\leq \ell\leq 1$. But we have
already obtained the general dimension of the primary field $\s_1$ in (\belun). With $p=5$ and $c=-3/5$, this
yields $h_1= -1/20$, which  identifies $\s_1$  to $\phi_{1,2}$. Its first descendant is 
$$\phi_{-\frac34+\frac12}|\s_1\R = \phi_{-\frac14}|\s_1\R \simeq |\phi_{1,3}\R \;,\eq$$
of dimension $1/5$.
With $\s_0\simeq \phi_{1,1}$ and $$\phi_{-\frac34}|\s_0\R \simeq |\phi_{1,4}\R\;,\eq$$
with dimension $3/4$,
the spectrum of the $\M(3,5)$ model is completely recovered.

Let us briefly comment on the origin of the operator $\S$ in the context of the graded parafermionic models, with coset
representation  $\osp(1,2)_k/\uh(1)$ [\CRS]. The $\M(3,5)$ model corresponds to $k=1$ [\JM]. Let $\psi_{\frac12}$
be the fundamental parafermionic field of dimension $1-1/4k$, which satisfies $(\psi_{\frac12})^{2k}\sim I$. 
Denote by
$\psi_1$ the parafermion of dimension $1-1/k$. We have $(\psi_{\frac12})^{2}\sim \psi_1$. For $k=1$  however,
$\psi_1\sim I$. But this is true up to a zero mode. Indeed, if we denote by $B$ and $A$ the respective modes of
$\psi_{\frac12}$ and $\psi_1$, then we have
$$B_{\frac14}A_{-1}|0\R = 0 = [A_{0}B_{-\frac34}+ B_{-\frac34}A_{0}]|0\R \;.\eq$$
We thus recover (\testa) with $B_{-\frac34}\simeq \phi_{-\frac34}$ and $A_0\simeq \S
$.

\subsec{The $\M(3,7)$ model }

In the present case, the central charge
is readily computed from
$$\underbrace{\phi_{-\frac14}\phi_{\frac14}}_{{\rm II}_{0,-1,1}}\phi_{-\frac54}|0\R=
\phi_{-\frac14} \underbrace{\phi_{\frac14}\phi_{-\frac54} }_{{\rm II}_{-1,0,0}}|0\R\;,\eq$$
leading to
$$-{1\over 16c} (7 c+25)\phi_{-\frac54}|0\R = 0 \;,\eq$$
with solution $c=-25/7$. 

The most interesting aspect to consider here, compared to the previous case, is the determination of the spectrum. In the $\M(3,5)$ case, we had two primary field: $\s_0=I$ and $\s_1$, whose dimension are directly determined by the commutation relations.  With $p=7$, we have three primary fields: $\s_0, \s_1$ and $\s_2$. Again the dimension of $\s_1$
results from (\gencb): $h_1= -5/28 \, (=h_{1,2})$. Here the difficulty lies in the determination of $h_2$, the
dimension of $\s_2$, which does not follow from a direct
application of the generalized commutation relations on $|\s_2\R$ . This dimension can
be fixed by associativity, however. For instance, by comparing :
$$\underbrace{\phi_{-\frac14}\phi_{\frac14}}_{{\rm II}_{-1,-2,3}}\phi_{-\frac14}|\s_2\R=
\phi_{-\frac14} \underbrace{\phi_{\frac14}\phi_{-\frac14} }_{{\rm II}_{-1,-1,2}}|\s_2\R\;,\eq$$
we get
$$\left\{{1\over 16} -{5\over 4c}\left(h_2+\frac14\right)\right\} \phi_{-\frac14}|\s_2\R= {5h_2\over
2c}\phi_{-\frac14}|\s_2\R \;, \eq$$
with solution
$h_2=-1/7$. We have thus
$$\eqalignT{
 &|\s_0\R\simeq |\phi_{1,1}\R\qquad &|\s_1\R\simeq |\phi_{1,2}\R\qquad &|\s_2\R\simeq  |\phi_{1,3}\R \cr
 &\phi_{-\frac54}|\s_0\R\simeq  |\phi_{1,6}\R\qquad &\phi_{-\frac34}|\s_1\R\simeq  |\phi_{1,5}\R\qquad
&\phi_{-\frac14}|\s_2\R\simeq  |\phi_{1,4}\R \;,\cr}\eq$$
and this complete the analysis of the $\M(3,7)$ model.

\newsec{The $\M(3,p)$ quasi-particle basis }

\subsec{The general form of the basis}

Let us now turn to a description of the basis of states for the $\M(3,p)$ models, with $p$ of both
parities. 

In the highest-weight module of $|\s_\ell\R$, the different states in the $N$-particle sector are of the form:
$$ \phi_{-h+\frac\ell2+\frac{(N-1)}2-n_1} \cdots  \phi_{-h+\frac\ell2+\frac{(N-i)}{2}-n_{i}} \cdots
\phi_{-h+\frac\ell2+\frac12-n_{N-1}}\, \phi_{-h+\frac\ell2-n_N}\, |\s_\ell\R\;,
\eqlabel\stru$$
with some constraints on the $n_i$'s to be specified below.
Note the cumulative contribution of the $\phi$ charge.
The highest-weight module is obtained by summing over all possible particle-sector  
$N$. The indices $n_i$ are integers for $p$ odd and alternate between integer and half-integer values when $p$ is even:
$$n_{N-2i}\in \Z \; , \qquad n_{N-2i-1}\in \Z+{p-1\over 2}\;. \eq$$
It is convenient to represent the string (\stru)  by a sequence whose $N$ entries (in the $N$-particle sector) are minus the modes $n_i$'s, i.e.,
$$  \phi_{-h+\frac\ell2+\frac{(N-1)}2-n_1} \cdots
\phi_{-h+\frac\ell2+\frac12-n_{N-1}}\, \phi_{-h+\frac\ell2-n_N} \quad \lra\quad (n_1,\cdots, n_{N-1}, n_N)\;,\eq$$
We will define our quasi-particle basis in terms of a filling process on a ground state. As argued below,
this ground state is
 described by the sequence
$$(\cdots 6,\, -r+5,\, 4,\, -r+3,\, 2,\, -r+1,\, 0)\;. \eqlabel\grosta$$
where we have introduced the notation
$$r={p-5\over 2} \;.\eq$$
Notice the increase of 2 units at distance 2 (from left to right).
The filling process amounts to add states corresponding to ordinary partitions on this ground state and sum over all
particle sectors.
The resulting sequences are not genuine partitions but merely `jagged partitions' [\JM, \ref{L. B\'egin, J.-F. Fortin, P. Jacob and P. Mathieu, Nucl. Phys {\bf B659} (2003) 365.}\refname\Beg,
\ref{J.-F. Fortin, P. Jacob and P. Mathieu, 
{\it Jagged partitions}, Ramanujan J. , to appear, math.CO/0310079.}\refname\FJMb] which satisfy
$$
n_i\geq n_{i+1}-r+1\;, \qquad 
n_i\geq n_{i+2}+2\;,\qquad  n_N\geq 0 \;. \eqlabel\jag$$
These conditions are directly read off (\grosta).  Note that for $r=0$ ($p=5$), 
these are standard partitions with distinct parts.  For
$r=1$ ($p=7$), (\jag) describes  standard partitions with a  difference 2 condition between parts separated by the distance 2. In the generic case $r>1$,  there is a
possible increase of $r-1$ between the part $n_{N-2i-1}$ and its right nearest neighbour  $n_{N-2i}$.   Note that the difference 2 at distance 2 implies a further possible increase of $r+1$ between  $n_{N-2i}$ and  $n_{N-2i+1}$ (again this is a direct consequence of (\grosta):  for $i=1$, $r+1$ is the difference between $2$ and $-r+1$). For instance, with  $p=14$, so that $r=9/2$, 
our candidate-basis in the vacuum module  is built on the ground state $(\cdots 6,\, 1/2,\, 4,\, -3/2,\, 2,\, -7/2,\, 0)$, that is:
 $$\cdots \phi_{-3+3-{\bf 6}}\, \phi_{-3+\frac52- {\bf \frac12}}\,  \phi_{-3+2-{\bf 4}}\,  \phi_{-3+\frac32+{\bf \frac32}}\,  \phi_{-3+1-{\bf 2}}\,  \phi_{-3+\frac12+{\bf \frac72}}\,  \phi_{-3-{\bf 0}}|0\R = \cdots \phi_{-6}\,  \phi_{-1}\,  \phi_{-5} \, \phi_{0}\, \phi_{-4}\, \phi_{1}\phi_{-3}\, |0\R\;. \eq$$
 (Note in particular that $\phi_{1}\phi_{-3}|0\R\propto L_{-2}|0\R$).\foot{Recall that $\M(3,14) \simeq W_3(3,7)$ model, so that $\phi=W$ in this context. Whether this basis can be lifted to a basis for the whole class of $W_3(3,p)$ models remains to be seen, however.}

Arguments supporting (\jag) are presented in the following subsection.

The characterization of the basis of states is not quite complete since yet there is no way of distinguishing the different highest-weight modules. Indeed, for $\ell\geq 2$, a  further restriction has to be
imposed on the parts. The origin of this  sort of boundary condition is simply that the lowest state in the 2-particle sector of the $\,
|\s_\ell\R$ module must be of the form
$\phi_{-h+\frac12+\frac\ell2-m_0} \,\phi_{-h+\frac\ell2}  \, |\s_\ell\R$ 
of dimension $2h-\ell-1/2+m_0$. But this state has to be proportional to $L_{-1}\, |\s_\ell\R$, which forces
$m_0= \ell-r $.
Since $m_0$ is the lowest value that $n_{N-1}$ can take, we have
$$n_{N-1}\geq \ell-r \;. \eqlabel\bdry$$
Our hypothesis would be that this is the whole set of boundary conditions. This is conformed by state counting at low levels.

 Let us illustrate the conditions (\jag)  as well as the boundary condition (\bdry) by listing the states of the $\ell=4$ module of the $\M(3,11)$ model at the  first levels. We do this by writing the corresponding sequences $(n_1,\cdots ,n_N)$.  We also  restrict ourself to the Virasoro module $| \phi_{1,5}\R$ which means that we only consider states that contain an even number of $\phi$ modes. The boundary condition (\bdry) requires $n_{N-1}\geq 1$. The descendant states up to level 6 are:
 $$\eqalignD{
 & 1\; : &\quad (1,0)\cr
  & 2\; : &\quad (2,0), \; (1,1)\cr
   & 3\; : &\quad (3,0),\; (2,1)\cr
    & 4\; : &\quad (4,0),\; (3,1),\; (2,2),\; (1,3),\; (3,2,1,0)\cr
    & 5\; : &\quad (5,0),\; (4,1),\; (3,2),\; (2,3),\; (4,2,1,0),\; (3,3,1,0)\cr
     & 6\; : &\quad (6,0),\; (5,1),\; (4,2),\; (3,3),\; (2,4),\;(5,2,1,0),\;(4,3,1,0),\;(3,4,1,0),\;(4,2,2,0),\;(3,3,1,0)\;.\cr}\eq$$
     For instance, the two sequences $(2,4)$  and $(5,2,1,0)$ correspond to the states
     $$ (2,4) : \phi_{-\frac14+\frac12-{\bf 2}}\, \phi_{-\frac14-{\bf4}}|\sigma_4\R\;,  \qquad (5,2,1,0): \phi_{-\frac14+\frac32-{\bf5}}\, \phi_{-\frac14+1-{\bf 2}}\, \phi_{-\frac14+\frac12-{\bf 1}}\, \phi_{-\frac14-{\bf 0}} |\sigma_4\R \;, \eq$$
     which indeed  both have  level 6.  The first state containing 6 modes $\phi$ occurs at level 9 and it is associated to the sequence $(5,4,3,2,1,0)$.

\subsec{The rationale for the condition (\jag)}

The aim of this section is to justify the basis (\jag) from conformal field theory.
Our argument is, to a large extend, maintained at a   sketchy level but  we expect that the main points can be recovered by a more rigorous 
analysis.

To investigate the structure of the $\phi$-type quasi-particle basis, we  consider the counting of independent states by
treating successively the different particle sectors (recall that the
`particle sector' is the
number of $\phi$-modes acting on the highest-weight state). In the first stage of the analysis, we only invoke the generic features of the $\phi$-algebra. Only in the later steps do we require this to be also a Virasoro minimal model.

We will stick
to the vacuum module
($\ell=0)$ for simplicity. It will prove convenient to rewrite the states
under the form
$$
\cdots  \phi_{h-b_{N-3}}\,\phi_{-h-a_{N-2}}\,\phi_{h-b_{N-1}}\,
  \phi_{-h-a_N}\, | 0\R\;.
 \eqlabel\struTa$$
Observe the specific choice made for the sign in
front of
$h$ within the modes which is designed to facilitate the use of the
commutation relations. In  this notation, the cumulative charge of
the $\phi$-mode is absorbed into the $a_i$ and $b_i$ labels.

In the one-particle sector, the states are of the form $\phi_{-h-n}\,
|0\R$. The only constraint is the
highest-weight condition (\vacuhw) which forces $n\geq 0$.

Consider next the two-particle sector. The
basic constraint here comes from the commutation relation (\genca). This
relation certainly continues to hold true also for the generic version of
the algebra under consideration, which does not affect the leading term of the OPE.
\foot{We stress that this
commutation relation holds for both parities of $p$ if we assume that
$\S
=I$ for $p$ even (in which case the infinite sum truncates to a finite
one). More generally, it holds for all value
of $p'$ with the  understanding that $\phi= \phi_{p'-1,1}$ and
$h=h_{p'-1,1}$.}  It is rather immediate to see
that all the states $\phi_{h-m}\,\phi_{-h-n}\, |0\R$  which do not satisfy
$m\geq n+2$ can be reexpressed in terms of those that do satisfy this
constraint.

Note that  $\phi_{h-m}\,\phi_{-h-n}\, |0\R$ with $m\geq n+2$ is equivalent
to  $\phi_{-h+1/2-n_1}\,\phi_{-h-n_2}\, |0\R$ with $n_1\geq n_2-r+1$.
This analysis of the two-particle sector can be directly transposed to a
sequence of two adjacent modes within  the bulk of a string of
$\phi$-modes, where the condition takes the form $n_i\geq n_{i+1}-r+1$.

So far, we have succeeded in explaining the jagged nature of the
sequences of $n_i$'s associated to the string of modes in descendant
states.  Next, we have to consider the independent states in the three-particle
sector. We consider states of the form
$$
\phi_{-h-n'}\phi_{h-m}\phi_{-h-n}|0\R \;, 
\eqlabel\phiThree
$$
and look for a constraint relating $n'$ to $n$.  At this point, we move away 
from the study of the most general $\phi$-type free basis and take into account the
constraints coming from the fact that the models we consider are also Virasoro minimal models.
As a result, every regular field appearing in the OPE $\phi (z) \phi (w)$
has to be rexpressible in terms of the energy-momentum tensor only.\foot{To be plain, this immersion amounts to a reduction of the allowed fields appearing in  the OPE $\phi (z) \phi (w)$. For instance, $(\partial GG) (w)$ is the first field so  removed for those superconformal models
that are also  Virasoro minimal models, namely for the $\M(3,8)$ and $\M(4,5)$ models.} This
means that when we compute the commutation relations between the $\phi$  modes, we can pick up regular terms of arbitrary order and still only
obtain Virasoro modes on the right-hand side of the commutation relations.  This in turns has the obvious consequence that
any state made of Virasoro modes can be expressed in terms of
$\phi$ modes (in even number) and vice-versa.  This is the advantage we take into
account in our next step.

We stress that our present objective is to determine some ordering condition on trilinear states by a recursive process. Given a target ordering condition, we need to show that states which do not satisfy this ordering condition can be expressed in terms of those which do satisfy it. With this in mind, we now introduce  a simplifying trick.

Let us return for a moment to the left-hand side of the commutation relations (\genca) and 
(\gencb).  Given the way these relations are derived, we see that the more terms we pick up in the OPE $\phi (z)
\phi(w)$, the higher is the gap between the left-most modes of the first sum and the right-most modes 
of the second sum.  Knowing that we can 
pick up any regular term (since they are all expressible as combinations of $T$), let us suppose that 
we select a regular term of sufficiently high order such that the second sum  contains only 
ordered states already considered so far in our analysis (that is, states that are more ordered than those under consideration at a given recursive step).  For the purpose of the present discussion, we can thus ignore the contribution of this second sum. That leads to a simplified version of the commutation relations (where the equality is to be understood as modulo terms previously considered):
$$
\sum_{i=0}^\y  c^{(i)} \phi_{-h-n'-i}\phi_{h-m+i}  = [L]_{-n'-m} + 
[LL]_{-n'-m}  + \cdots
\eqlabel\newCom
$$
where in this notation, the $c^{(i)}$ are unspecified constants and $[L \cdots L]_{-N}$ (with $n$ factors  of $L$)  represents a given linear combination of $n$  $L$ modes at level $N$. All the coefficients are fixed by the proper choice of commutation relations and the values of $n'$ and $m$.
 
Let us now assume at this point that the terms $n' < n-1$ in (\phiThree) can be reorganized 
in terms of those with $n'< n$.  Using (\newCom) in 
(\phiThree) for $n' = n-1$, we obtain:
$$
\eqalign{
 &  \phi_{-h-n+1}\phi_{h-m}\phi_{-h-n}|0\R \cr
&\qquad = \left[\left( [L]_{-n-m+1} + [LL]_{-n-m+1}+ \cdots \right)  + \sum_{i=1}^\y c^{(i)} \phi_{-h-n-i+1}\phi_{h-m+i}\right] \phi_{-h-n}|0\R \cr
&\qquad= a\phi_{-h-2n-m+1} |0\R + \phi_{-h-n} \left( [L]_{-n-m+1} + 
[LL]_{-n-m+1}+ \cdots \right) |0\R 
 + \sum_{i=1}^\y  c^{(i)} \phi_{-h-n-i+1}\phi_{h-m+i}\phi_{-h-n}|0\R \cr
&\qquad= a \phi_{-h-2n-m+1} |0\R + \phi_{-h-n} \sum_{i=0}^\y  c'^{(i)} 
\phi_{h-n-m+i+1}\phi_{-h-i}|0\R +
 \sum_{i=1}^\y  c^{(i)} \phi_{-h-n-i+1}\phi_{h-m+i}\phi_{-h-n}|0\R \cr
}
\eqlabel\long
$$
where $a$ is some constant. In the second equality, $\phi_{-h-n}$ has been commuted with the Virasoro modes, while in last one,  the Virasoro  modes  have been reexpressed in terms of the 
$\phi$ modes.  

We see that most  of the   $\phi$-trilinear states  can be now written in the more 
ordered form $n' \geq n$.  If we get rid of the states already 
considered so far, we see that our 
initial state can be expressed in terms of 
the remaining non-ordered terms as:
$$
 \phi_{-h-n+1}\phi_{h-m}\phi_{-h-n}|0\R 
= [\phi_{-h-n}\phi_{h-m+2}\phi_{-h-n-1} + 
\phi_{-h-n-1}\phi_{h-m+4}\phi_{-h-n-2} + \cdots ]|0\R\; .
\eq
$$
The sum has to stop at some point as we can use the reordering 
of the bilinear terms in $\phi$.  Now we can repeat this for these new non-ordered states
 until we ultimately run out of such non-ordered terms.  In other words, by starting with a given 
$n'=n-i$, a recursive process  allows us to obtain states of the form $n' > n-i$ for $i>0$.

If we try to apply the same trick to eliminate states of 
the form $n' > n$, we end up with relations linking the $n'> n$ states 
to $n' < n$ states. By consistency, we expect these relations to be 
the same ones we have first obtained for $n'<n$.

Finally, if we consider the states with $n' = n$, we now notice that these states with $n' = n$ reappear along the derivation. Without 
explicitly calculating every coefficients in front of the states, we 
do not know whether some of the $n' = n$ states can be eliminated or not.

We can thus naturally expect that the conditions on the trilinear terms 
in $\phi$ will lie somewhere in between the conditions $n' \geq n$ 
and $n' > n$. Note that the above analysis can be applied to all $\M(p,p') $ models if we replace $\phi$ by $\phi_{p'-1,1}$. From low-level state-counting  checks, we can verify that the previous property  is indeed verified for any $\M(p',p)$ model with $p'\geq 3$ (which ensures that $\phi_{p'-1,1}\not=\phi_{1,1}$). The most restricted case corresponds to $p'=3$ because  the $\phi$ singular vector arises at the lowest possible level, which is 2.    This case lies exactly on the upper-bound constraint, that is, $n' > n$. In order to get the least restricted case, we have to examine the models for which  the singular vector appears as deeply as possible  in the module.  As $\phi_{p'-1,1}$ has its first singular vector at level $p'-1$, it corresponds to cases where $p'$ is large.  It appears that the trilinear terms in those cases are actually more restricted than $n' \geq n$, indicating that this lower-bound condition is not saturated.

We have just seen how the condition $n' \geq n$ had to be respected.  Now we will try to be slightly more explicit about the possible restrictions for $n'=n$.  Let us consider (\newCom) once again along with (\phiThree).  Depending on the value of $m$, we do not always need to pick up the  same regular term in order to eliminate the second sum in the commutation relations.  The higher the value of $m$, the greater is the number of possibilities we have to write the commutation relations.  Using all these different choices of commutation relations, we get a number of linearly independent relations taking all the form of the first equality in (\long), 
with the coefficients before each terms differing  from one choice to the other.
If one could keep track of all the conditions coming form these different equalities, this could lead us to two possible outcomes.  On the one hand, the result might be some intermediate `basis' between the spanning set of states and the complete sought-for basis, upon which we would have to apply the restrictions coming from the removal of the first null-field
in $\phi$. On the other hand, the result could actually take care of all possible constraints. If this second possibility is the actual one, it would mean that the immersion of the $\phi$-extended conformal field theories into the $\M(p',p)$ models fixes completely the Virasoro singular-vector structure.

We end up this section by displaying a sample computation supporting the later alternative. The example to be considered is the one for which $\phi$ has dimension 3/2. The $\phi$-algebra is thus a superconformal algebra. Considered also as a Virasoro minimal model (which embodies a truncation of the space of states), this is associative for two values of $c$, corresponding to the $\M(3,8)$ and  $\M(4,5)$ models. We will show that in the former case, the state
$\phi_{-\frac32}\phi_{-\frac12}\phi_{-\frac32}|0\R $ can be eliminated.


 The anticommutation relation for $\phi=G$ takes the form
$$\{ G_{n+\frac12}, G_{m-\frac12} \}= \frac{n(n+1)}2
\delta_{n+m,0}+\frac{c}3 L_{m+n} \;. \eqlabel\gcom$$
This is obtained by considering only the singular terms in the OPE
$G(z)G(w)$. But if, instead, we pick up the contribution  of more terms,
in particular, up to and including the level-two descendants of $T$, we
then get
$$
\eqalign{
&\sum_{t\geq 0}C_{-2}^{(t)}\, [G_{n+\frac12-t}G_{m-\frac12+t}  +
G_{m-\frac52-t}G_{n+\frac52+t}] = \frac{(n+3)!}{24(n-1)! }\delta_{n+m,0}\cr
& \qquad +\left[\frac{3(n+3)(-m)}{2c}+\beta_1(n+m+3)(n+m+2)\right]L_{n+m} \
 +  \beta_2\sum_{t\geq0} \,
[L_{-2-t}L_{n+m+2+t}+L_{n+m+1-t}L_{-1+t}]\;,\cr}\eqlabel\gcoma$$
with $\beta_1$ and $\beta_2$ being  the coefficients of $T''$ and $(TT)$
respectively. These constants are fixed by associativity to the values
$$ \beta_1= {9(c+1)\over 4c(22+5c)}\qquad {\rm and}\qquad \beta_2=
{51\over 2c(22+5c)}\;.\eq$$
Further calculations show that the central charge must be restricted to
the two values $ 7/10$ and $-21/4$ as previously said.
Using (\gcom), we can write
$$G_{-\frac32}G_{-\frac12}G_{-\frac32}=  {3\over c}\, G_{-\frac32} L_{-2}
= {3\over c}\left [G_{-\frac72}+ L_{-2}G_{-\frac32}\right] \;. \eq$$
Using now the  commutation relations (\gcoma) with $n=m=-1$ in order to
get an expression for $L_{-2}$ (as given by the first term on the right-hand side of
(\gcoma)) acting on $G_{-3/2}$, we  find that
$$L_{-2}G_{-\frac32} |0\R = {3\over c}\left[ \left(-3\beta_2-{6\over c} \right)
L_{-2}G_{-\frac32} +
\left(-2\beta_2+5-{3\over 2 c}\right)G_{-\frac72}\right] |0\R \;.\eq$$
For $c=7/10$, the second coefficient on the right-hand side vanishes  while the first
one reduces to 1; in other words, we end up with the identity
$L_{-2}G_{-\frac32}|0\R = L_{-2}G_{-\frac32}|0\R$. Therefore, in that case, there
is no relation between $L_{-2}G_{-\frac32}|0\R $ and $G_{-\frac72}|0\R $. But for
the other allowed value of $c$, which corresponds to that of the $\M(3,8)$
model, there is one such relation. (Actually, we have recovered here the
expression for the $\phi_{2,1}$ singular vector.)
 It implies  that, in this precise case, we can eliminate the state
$G_{-\frac32}G_{-\frac12}G_{-\frac32}|0\R $.

 Higher order terms can be treated along these lines. 
 But clearly, going deeper in the
modules requires the computation of more and more terms in
$\phi(z)\phi(w)$. These computations are thus rather complicated, in
addition to be  model-dependent. But they provide independent
verifications of the stated conditions (\jag).


\newsec{The $\M(3,p)$  fermionic-type  characters}

Our main  assumption is that (\jag) provides a basis. This has been supported   by heuristic  considerations, some explicit computations and the comparison with known bases for small values of $p$. However, establishing (\jag) rigorously is a hard mathematical problem. We circumvent this by showing  that these conditions do indeed lead us to the expected characters.
 More precisely, we demonstrate here  that the enumeration of states subject to the conditions (\jag)  together with the boundary condition (\bdry), reproduces the known expressions for the $\M(3,p)$ characters in their fermionic form.

In view of enumerating all the states in a given module (i.e., constructing its character), it is convenient to
transform the ground state into one for which the parts do not increase from left to right.  Let us then add to
the ground state (\grosta) the staircase of $(r-1)$-height step:
$$(\cdots,\,  5r-4,\, 4r-3,\, 3r-2,\, 2r-1,\, r,\, 1) \; . \eqlabel\stair$$
The shifted  ground state is thus\foot{We stress that this shifting is simply a relabeling of the ground state. Note also if the ground state (\grosta)
involves both integers and half-integers for $p$ even, the shifting process generates only integer parts since $2r$ is always
integer.}
$$(\cdots, 4r+1,\, 4r+1,\, 2r+1,\, 2r+1,\, 1,\, 1)\;. \eqlabel\shif$$ 
Partitions defined on this shifted ground state can be characterized as follows.  These are partitions
$(\la_1,\la_2,\cdots,\la_N)$ of length
$N$,
$$\la_i\geq
\la_{i+1}\; , \; \qquad \la_N\geq 1 \;, \eq$$  
satisfying the supplementary condition
$$\la_i\geq \la_{i+2}+2r \;. \eqlabel\drco$$
These
conditions follow from (\jag) and
$$\la_i= n_i+(N-i) (r-1) +1\;. \eq$$
In other words, parts in $(\la_1,\cdots,\la_N)$ that are separated by the distance 2 must then differ by at least  a $2r$, with $2r=p-5$.\foot{If, instead, we subtract from the ground state
(\grosta) the staircase
$(\cdots, -r+2,  -r+1,-r)$, it becomes $(\cdots 0,r,0,r,0,r)$. This 
is the ground state of jagged partitions of type $0r$
in the terminology of [\FJMb].}. To these conditions, we need to add the boundary condition:
$$\la_{N-1} \geq \ell   \;. \eqlabel\bryco$$ 


Let $p_{r,\ell}(w,N)$ be the number of  partitions
 of length 
$N$ and weight
$w$ (that is, $w=\sum_i\la_i$) satisfying (\drco) and (\bryco). Denote the corresponding generating function by
$$G_{r,\ell}(z,q)
= \sum_{w,N\geq 0} \; p_{r,\ell}(w,N)\, q^w z^N \;. \eqlabel\gfr$$
For $0\leq \ell\leq k$, this function can be obtained in closed form as a $k$-multiple sum:
$$G_{r,\ell}(z,q)
= \sum_{m_1,m_2,\cdots m_k\geq 0} {q^{{\bf m}\, {\tB} \, {\bf m} + \tC\cdot {\bf m}}
\; z^{2(m_1+\cdots+ m_{k-1}) + m_k}\over (q)_{m_1}\cdots (q)_{m_k}}\;,\eqlabel\genf$$
where $k$ is related to $p$ by
$$k=\left[p\over 3\right]\;, \eq$$
(where $[x]$ stands for the integer part of $x$)
and it is understood that 
$$ {\bf m}\, {\tB} \, {\bf m}= \sum_{i,j=1}^k m_i \, \tB_{ij}\,  m_j, \qquad \tC\cdot {\bf m}=
\sum_{i=1}^k {\tC}_i\,m_i\;.
\eq$$
The $k\times k $  symmetric  matrix $\tB$ reads
$$\tB= \pmatrix {2r&2r&\cdots &2r &{r}\cr 2r&2r+1&\cdots &2r+1 &r+\frac12\cr\cdots&\cdots
&\cdots&\cdots &\cdots\cr 2r&2r+1 &\cdots &2r+k-2&r-1+\frac{k}2\cr
 {r}&r+\frac12 &\cdots
&r-1+\frac{k}2& k-1\cr}\;,\eqlabel\deB$$
while the vector $\tC$ takes the form
$$\tC_j=-2r+j+1+{\rm max}\; (\ell-j,0) \quad {\rm for}\quad j<k \qquad {\rm and} \qquad \tC_k=-k+2\;. \eqlabel\deCL$$
Finally, in the denominator of (\genf),  we made use of the notation
 $$(q)_n= \prod_{i=1}^n(1-q^i)\;. \eq$$


The generating function (\genf) for $\ell=0$ (which turns out to hold also for $\ell=1$) has been found in
[\FJM]
(cf. Theorem 5.14 with
$M_i,N\rw \y$).  For $\ell\not=0$ but $\ell\leq k$, the boundary condition (\bryco) induces a modification by terms linear in the $m_i$'s and these
are easily fixed by looking at the lowest partitions in low ($N=1,2$)  particle sectors. This is taken care by the  second term in $C_j$, as demonstrated in appendix B.  The latter is a  verification proof  and it breaks for $\ell>k$. A further analysis of these boundary terms is presented in appendix C. It is shown there that $G_{r,\ell}$ can be recovered from $G_{r,0}$ recursively, for all $1\leq \ell\leq p/2-1$. However, it appears that it is only for $1\leq \ell \leq k$ that the $G_{r,\ell}$ can be reconstructed in closed form, as a single fermionic multisum. 

Our goal is to count partitions built on the ground state (\grosta)  subject to (\bdry) and weight them by their proper
conformal dimension, constructing thereby the character of the $|\s_\ell\R$ module.
 We thus start with $G_{r,\ell}(z,q)
$ and enforce $z^N$ to be equal to a certain power of $q$ adjusted 
in order to: 1- undo the ground state shifting by taking out the staircase 
contribution (\stair), whose weight is denoted $w_{{\rm stair}}$;  and 2-
add the fractional part, with  weight $w_{{\rm frac}}$, to correct for the fact that (\grosta) 
does not take into account
the fractional part of the $\phi$-modes. 
These numbers are easily computed.
On the one hand, the staircase $((N-1)r-(N-2),\cdots,2r-1,r,1)$ has
weight 
$$w_{{\rm stair}}= {N\over 2}\left[(N-1)(r-1)+2\right]\;. \eq$$
On the other hand, the fractional part has the following dimension
$$w_{{\rm frac}}= N\left[h-{\ell\over 2}- {(N-1)\over 4}\right]={ÊN\over 4} \left(2r+4-2\ell-N\right) \;,\eq$$
where the third  term in the square bracket comes from the cumulative contribution of the $\phi$ charges.
We thus replace $z^N$ in (\genf)  by $q^{w_{{\rm frac}}-w_{{\rm stair}}}$, where
$$N= 2(m_1+\cdots + m_{k-1}) + m_k\;. \eq$$ This leads to the following expression for the character $\hat{\chi}_\ell$ 
(normalized such that its leading $q$ power is $q^0$, hence the hat),
$$\hat{\chi}_\ell(q)= \sum_{m_1,m_2,\cdots m_k\geq 0} {q^{{\bf m}\, {B} \, {\bf m} + C\cdot {\bf m}}
\over (q)_{m_1}\cdots (q)_{m_k}}\;, \eqlabel\genfa$$
where, with $1\leq i,j\leq k-1$
$$ B_{ij}= {\rm min} (i,j) \; , \qquad 
 B_{jk}=B_{kj} = j/2 \; , \qquad  B_{kk}= {k+1-\e\over 4}\;, \eq$$
and  $C$ reads
$$C_j= {\rm  max}\; (j-\ell,0)\;, \qquad C_k= {k-1+\e -\ell\over 2}\;, \eq$$
with $\e=0,1$ defined by
$$p= 3k+1+\e\;. \eqlabel\defiu$$ In terms of the Virasoro characters, $\hat{\chi}_\ell$ decomposes as follows
$$\hat{\chi}_{s-1}(q)= q^{-h_{1,s}+c/24} \left[\chi^{{\rm Vir}}_{1,s}(q)+ q^{h_{1,p-s}-h_{1,s}}\chi_{1,p-s}^{{\rm
Vir}}(q)\right]\;.\eq$$ Recall that in our construction, the Virasoro primary field $\phi_{1,p-s}$ (for $p>2s$) is a
$\phi$-descendant of
$\phi_{1,s}$. The two Virasoro characters can be separated by the parity of $N$, which is the same as that of $m_k$: with $m_k$
even (odd) , we obtain $\chi_{1,s}$ ($\chi_{1,p-s}$ respectively). 

We recover thus the fermionic sums given in [\By] in a form similar to that displayed here.  Their complete proof is presented
in [\Wel] and their reexpression in the above form is worked out in [\FFW].


\newsec{Conclusion}

In this work, we have considered the reformulation of the minimal models $\M(3,p)$  in terms of the algebra spanned by $\phi\equiv \phi_{2,1}$ and defined by the OPE (\opep). The structure of this algebra differs somewhat according to the parity of $p$: for $p$ even, $2h\in \Z_+$ while for $p $ odd, $h \in \Z_+\pm 1/4$.  In the latter case, the associativity analysis forces the introduction of a Witten-type operator anticommuting with $\phi$. A similar operator (but presented differently) has been found in the spinon formulation of the $\suh(2)_1$ model [\BLS]. It seems to characterize non-local algebras with generators of dimension $h \in \Z_+\pm 1/4$

The Hilbert spaces (highest-weight states and their descendants) have been completely described in terms of the $\phi$-algebra. In particular, the modules are described by the successive action of the lowering $\phi$-modes subject to specific constraints. In the $N$-particle sector, with strings of lowering modes written in the form (\struT), these constraints are given in (\jagG).    
 The  highest-weight states themselves are distinguished by the integer $\ell$ whose range is $0\leq \ell\leq p/2-1$. Moreover, the $\ell$-dependence of the descendant states is fully captured by the condition $n_{N-1}\geq \ell-r$.
The obtained basis agrees with those previously found for $p=4,\, 5$ and $8$ and the one derived in [\FJM, \FJMMT] for $\ell=0$.

 
In absence of a   complete  argumentation underlying the derivation of this basis, our considerations  have been supplemented by general arguments and explicit computations relying on the  observation that the fine structure of the defining $\M(3,p)$ OPE $\phi(z)\phi(w)$ encodes the complete information on the models, including its quasi-particle basis.
Note that this analysis does not
 mimic
that of the $\M(2,p)$ and ${\cal SM}(2,4\kappa)$ cases. In these cases,  the spanning  basis of states is first obtained and  a set of restrictions, 
arising from the identity  null field, is then imposed. The spanning set of states  for the $\phi$-algebra, on which one could impose constraints such as the level-two $\phi$ null field, has not been found yet.

The simplest way of verifying the correctness of the basis of states is to derive the character of the irreducible module of $|\s_\ell\R$. This  is obtained by enumerating  all these states (\struT)-(\jagG) and summing over  all values of $N$.  The character is explicitly given by (\genfa) when $0\leq \ell\leq  [p/3]$.  This agrees with the known fermionic form of these characters [\By, \FFW].  Although for $[p/3] < \ell\leq p/2-1$, the character has not been found in closed form, it is shown in appendix C how it can be obtained recursively form $G_{r,0}$. But we stress that the validity of the boundary term in this range has been tested by writting explicitly the states at the first few levels of various modules and comparing  their enumeration with that given by the usual bosonic formula.

The $\M(3,p)$ models, due to the presence of an
extra symmetry generator,
 are somewhat similar to the superconformal models. For the special class of ${\cal
SM}(2,4\kappa)$ models, we can either choose to write the quasi-particle basis either  in terms
of the modes of $G$  together with  the
Virasoro modes or solely in terms of the $G$ modes [\FJMa]. This suggests that one could look for an
 alternative quasi-particle basis for the $\M(3,p) $ models, this one  formulated in term of 
an ordered set of Virasoro modes acting on an ordered set of ${\phi} $ modes as
 $$
 L_{-n_1} \cdots L_{-n_k} {\phi}_{-m_1}  \cdots{\phi}_{-m_{k'}} |0 \rangle  \eqlabel\mixed$$
with
some constraints on the numbers $n_i$ and $m_i$. A basis of that type  has indeed been found;  this result will be presented elsewhere [\ref{P. Jacob and P. Mathieu, {\it A new quasi-particle basis for  the $M(3,p)$ models}, in preparation.}].

Another natural axis for extension follows from  the observation that an analysis similar to the present one should be applicable to all extended algebras having  the essential simplifying property of being single-channel.

Let us conclude by emphasizing the fact  that there is a relatively small number of  conformal field theories for which the fermionic characters are described 
in terms of a basis derived by
intrinsic conformal field theoretical methods. We have already mentioned that this is so for the $\M(2,p)$ minimal
models [\FNO] together with their superconformal analogues, the ${\cal
SM}(2,4\kappa)$ models [\FJMa]. But there are few other examples, like  the $\su(2)_k$ models [\BPS,\BLS], particular higher-rank WZW models [\ref{ P. Bouwknegt. and K. 
Schoutens, Nucl. Phys. 
{\bf B 547} (1999) 501. }\refname\bouschou, \ref{G. Georgiev, 
J. Pure Appl. Algebra {\bf 112} (1996) 247.}], the parafermionic models 
[\JMb],  their graded version [\JM, \Beg] and some higher-rank formulations [\ref{G. Georgiev,  {\it  Combinatorial constructions of modules for infinite-dimensional Lie algebras, II. Parafermionic space}, q-alg/9504024.}\refname\Geo].  The present  analysis is  a step toward the addition of a further example to this list, the
$\M(3,p)$ minimal models.

\appendix{A}{Associativity and Jacobi identities}

The associativity conditions for the symmetry generators of an extended conformal algebra are sometimes loosely viewed as being equivalent to the Jacobi identities for the
mode-generators. For fields with (half-)integer dimension that associativity implies the Jacobi
identity can indeed be derived in a rather direct way.\foot{Given three integer-dimension operators $A,\,B,\, C$  with ordinary commutation relations, let us consider the action of
$A_nB_mC_p$ on  an arbitrary state $|h\R$. The associativity requirement retranscribed at
the level of modes implies  that this state can be evaluated by commuting the first two terms or the last
two ones without affecting the result. This in turn implies the usual form of the Jacobi identity as we now
show.  Commuting $A$ and $B$ and reexpressing the result in terms of the state $C_pB_mA_n|h\R$ yields
$$\left( \,[[A_n,B_m],C_p]+ C_p[A_n,B_m]+B_m[A_n,C_p]+[B_m,C_p]A_n+  C_pB_mA_n\, \right)|h\R\;. $$
Commuting $B$ and $C$ and singling out again the state $C_pB_mA_n|h\R$ gives
$$\left( \,[A_n,[B_m,C_p]]+ [B_m,C_p]A_n +[A_n,C_p]B_m +C_p [A_n,B_m]+  C_pB_mA_n\, \right)|h\R\;. $$
The comparison of these two expressions implies the identity
$$\left( \,[[A_n,B_m],C_p]+ [[C_p,A_n],B_m]+[[B_n,C_p],A_n]\, \right)|h\R=0\;. $$
This is the Jacobi identity; more precisely, this is the Jacobi identity modulo a singular vector
of $|h\R$. For fields with half-integer dimension, this is also
true but with an appropriate graded version of the Jacobi identity.}
But 
the reverse is not true. As the following considerations will illustrate, associativity contains more information than the mere 
Jacobi identities for the modes.

Consider a free fermion, whose OPE and mode decomposition (in the
 NS sector) read:
$$\psi(z)\psi(w)\sim{1\over z-w} \;, \qquad\qquad \psi(z)= \sum_{n\in\Z+1/2} b_n z^{-n-1/2}\;. \eq$$
 By evaluating the integral 
$${1\over (2\pi i)^2}\oint_{0} dw\oint_{w} dz\, {z^{n-1/2} w^{m-1/2} } \psi(z)\,
\psi(w)\;, \eqlabel\genin$$ 
 we find the usual anti-commutation relations:
$$\{ b_n, b_m\}= \delta_{m+n,0}\;. \eq$$ 
Considering this anticommutator together with the Virasoro commutation relations and
$$[L_n,b_m]=-\left({n\over 2}+m\right)b_{n+m}\;,\eq$$ it is simple to convince oneself that the central charge is not fixed by the
Jacobi identity. However, if we consider the following version  of the OPE
$$\psi(z)\psi(w)= {1\over z-w}+{z-w\over c} T(w)+\cdots\;, \eqlabel\flop$$
(given that the $\beta^{(2)}$ coefficient is $2h_\psi/c=1/c$) and the integral
 $${1\over (2\pi i)^2}\oint_{0} dw\oint_{w} dz\, {z^{n+1/2} w^{m+1/2} \over (z-w)^2} \psi(z)\,
\psi(w)\;, \eqlabel\genint$$ we get the following generalized relations:
$$\sum_{l\geq 0} l\; [ b_{n-1-l}b_{m+1+l}+b_{m-1-l}b_{n+1+l}]= {(n-1/2)(n+1/2)\over 2} \delta_{n+m,0}
+{1\over c} L_{n+m}\;, \eqlabel\fulcom$$
or equivalently (cf. [\ref{P. Jacob and P. Mathieu, Nucl. Phys.
{\bf B 587} (2000) 514.}\refname\JMa], App. C)
$$\sum_{l\geq 0} (l+1)\; [ b_{n-3/2-1}b_{m+3/2+l}+b_{m-1/2-1}b_{n+1/2+l}]= {n(n-1)\over 2}
\delta_{n+m,0} +{1\over c} L_{n+m}\;.\eqlabel\fulcom$$
These relations capture the expression of the energy-momentum of the free fermion in terms of
modes (but as a function of the yet-to-be-fixed central charge). In other words, the relation between $T$
and
$\psi$ is already coded in the OPE and the conformal invariance.\foot{This, of course, is a totally
standard statement since the OPE (\flop) can also be written as
$$\psi(z)\psi(w)= {1\over z-w}+(\psi(z)\psi(w))={1\over z-w}+ (z-w)(\d \psi
\psi)(w)+\cdots\;,$$ from which we read that $$T(w)= c (\d \psi
\psi)(w)= -c( \psi
\d \psi)(w)\;, $$ and this corresponds to the usual expression when $c=1/2$.} Note that the
four-point function
$$\L\psi_1(z_1)\psi_2(z_2)\psi_3(z_3)\psi_4(z_4)\R= {1\over z_{12}z_{34}} - {1\over z_{13}z_{24}}+ {1\over
z_{14}z_{23}}\;, \eq$$
which is evaluated solely from the knowledge of the singular terms (through meromorphicity, cf. [\ref{P. Jacob and P. Mathieu, J. Phys. A: Math. Gen.  {\bf 34} (2001) 10141.}]),
contains the information on the central charge when viewed in the light of the OPE (\flop). It is 
extracted  by evaluating the correlator in the limit
$z_1\rw z_2$.  This does not contradict the previous conclusion because the correlation function encompasses the whole content of the involved OPE.

 The  free-boson theory offers another simple illustration of the gap between the information obtained
from associativity and the Jacobi identities. The OPE
$$ i\d\vp(z)\, i\d\vp(w)= {1\over (z-w)^2}+{2\over c} T(w)+\cdots \;,\eq$$
and mode decomposition
$$ i\d\vp= \sum_n a_n
z^{-n-1}\;, \eq$$ lead to the standard mode commutation relation as
$$[a_n,a_m]={1\over (2\pi i)^2}\oint_0 dw\oint_wdz\, z^n w^m i\d \vp(z)\, i\d\vp(w)=
n\delta_{n+m,0}\;,\eqlabel\boco$$ 
for which the Jacobi identity is trivial. If we consider instead the integral
$${1\over (2\pi i)^2}\oint_{o} dw\oint_{w} dz\, {z^n w^{m+1} \over (z-w)} i\d \vp(z)\,
i\d\vp(w)\;, \eqlabel\bocom$$ we end up with a very different form of the commutation relations, namely,
$$\sum_{l\geq 0} [ a_{n-l-1}a_{m+1+l}+a_{m-l}a_{n+l} ]= {n(n-1)\over 2} \delta_{n+m,0} +{2\over
c}L_{n+m}\;. \eq$$ 
It is easily checked that $c$ is fixed to $1$ by associativity. This last expression is  thus seen to be the  mode expression of the
energy-momentum of the free boson, i.e., 
$$L_n= \sum_{l\in\Z}a_l a_{n-l} \;, \qquad (n\not=0)\qquad L_0= \sum_{l\geq 0}a_{-l} a_{l}
+\frac12 a_0^2\;. \eq$$
This  observation is of course true in general: picking up the energy-momentum as the single pole, yields directly the expression of the
Virasoro modes  in terms of the conserved-current ones.

These simple considerations show that the OPE contains more information than a
particular form of commutation relation derived from it. To recover the complete information
which is contained in the OPE, we need to consider the infinite family of commutation relations that follows
from evaluating (\genint) with $(z-w)^2\rw (z-w)^p$, for all values of $p$. In practice,  however, only few values
of
$p$ should be sufficient.


The associativity of the four-point functions $\L A B C D\R$ is equivalent to the statement that the state
$A_\ell B_m C_n|h\R$ where $h$ is the dimension of $D$, is independent of the way it is evaluated. 
Generically, the resulting constraints are independent of the state $|h\R$ (unless an equality is true
modulo a singular vector) and in practice it can be replaced by the vacuum. To formulate the mode version
of the associativity constraints, we introduce the notation
$$\underbrace{ A \, B}_{p}\, C\equiv [A,B]_p\, C + B\,A\,C\;, \eqlabel\nota$$
where $[A,B]_p$ stands for the commutator evaluated in terms of the  generalized commutation relations
that follow from evaluating the integral:
$${1\over (2\pi i)^2}\oint_{0} dw\oint_{w} dz\, {z^{n'} w^{m'} \over (z-w)^p} A(z)\,
B(w)\;, \eq$$
(the values of $n'$ and $m'$ being adapted to the choice of $p$ in order to recover a final commutator
in standard form -- cf.  (\genin) vs  (\genint) and (\boco) vs (\bocom)). Now, mode-associativity boils
down to the following conditions:
$$\underbrace{ A \, B}_{p}\, C =  \underbrace{ A \, B}_{q}\, C  = A\, \underbrace{ B \, C}_{p}\;.
\eqlabel\assmo$$
In principle, these conditions should be tested for all values of $p$ and $q$ and all combinations of modes.  However, in practice,  a small number 
of computations of this type are needed to fix the whole structure of the models under consideration.

As a simple illustration, let us show how we can fix the central charge of the free-fermion theory by
enforcing the mode-associativity of $b_{-1/2}b_{1/2}b_{-1/2}|0\R$. For this we compare  
$$ b_{-1/2} \underbrace{b_{1/2}\,b_{-1/2}}_{p=0}|0\R = b_{-1/2}|0\R\;, \eq$$
(the $p=0$ commutator being (\genin)) to 
$$ \underbrace{  b_{-1/2} \, b_{1/2}}_{p=2}\,b_{-1/2}|0\R ={1\over c} L_0 b_{-1/2}|0\R
= {1\over 2c} b_{-1/2}|0\R\;, \eq$$
(the $p=2$ commutator being (\genint)) to find that $c=1/2$. Note that in this case, we could also compute
the central charge by comparing the last result with
$$ b_{-1/2} \underbrace{b_{1/2}\,b_{-1/2}}_{p=2}|0\R = b_{-1/2}|0\R\;, \eq$$
(in the first case, we generate a term proportional to a Virasoro mode, hence containing $c$, and in the
second case the remaining contribution is the delta term, independent of $c$).


\def\Ol{{\displaystyle\Om_\geq^a}}
\def\Ola{{\displaystyle\Om_{\geq}^{a_1}}}
\def\Olaa{{\displaystyle\Om_{\geq}^{a_2}}}
\def\Olaaa{{\displaystyle\Om_{\geq}^{a_3}}}

\appendix{B}{Boundary terms in the generating function}

Our  goal is to derive the modification of the generating function $G_{r,0}(z,q)
$ (\genf), that counts the partitions $(\la_1,\cdots, \la_N)$ with $\la_i\geq \la_{i+2}+2r$ with $\la_N\geq 1$ [\FJM], which results from the further condition $\la_{N-1}\geq \ell$. 
We start with the assumption that the boundary terms are represented by linear factors in the exponent, i.e., are accounted by a correction of the form $q^{\sum_{j=1}^k m_j a_j}$.  Denote the modified form as $G_{r,\ell}(z,q)
$. Since the boundary condition is well localized i.e., it concerns only the second term at the right,  it suffices to consider the $N=1,2$ sectors to fix the $a_j$. Recall that $N= 2\sum_{j=1}^{k-1} m_j+m_k$. Therefore, $N=1$ corresponds to $m_k=1$ and all the others $m_j=0$. The generating function $G_{r,0}$ multiplied by $q^{a_k}$ reads
$$G_{r,\ell}(q,1)= {q^{1+a_k}\over (q)_1} \qquad (N=1)\;. \eq$$
But the partitions that are to be counted are simply those with a single part and their generating function is $q/(q)_1$. This fixes $a_k=0$.  

Consider next $N=2$ which requires either that a single mode $m_j =1$ for $1\leq j\leq k-1$ with all other modes zero or that $m_k=2$, again with all other modes vanishing. This results into
$$G_{r,\ell}(q,1)= {1\over (q)_1}\sum_{j=1}^{k-1}q^{2j+a_j} +{q^{2k}\over (q)_2} \qquad (N=2)\eqlabel\gftwo$$
In order to fix the $a_j$, we must determine the generating function enumerating  partitions with two parts $(\la_1,\la_2)$ and satisfying
$$\la_1\geq \la_2\geq 1\qquad {\rm and} \qquad\la_1\geq \ell\; .\eq$$This can be done by means of  the MacMahon method  [\ref{P.
MacMahon, {\it Combinatory analysis}, 2 vols (1917,1918), reprinted by Chelsea, third edition,
1984.}\refname\Mac] (cf. vol 2 Sect. VIII, chap. 1 and see also [\ref{
G.E. Andrews, R. Askey and R. Roy, {\it Special functions}, Encyclopedia of Mathematics and its applications {\bf 71},
Cambridge Univ. Press (1999).}], Sect.. 11.2), by  projecting the following expression
$${a_3^{-\ell} a_2^{-1}\over (1-a_1a_3 q)(1-a_2 q/a_1)}= \sum_{\la_1,\la_2\geq 0} a_1^{\la_1-\la_2}\, a_2^{\la_2-1}\, a_3^{\la_1-\ell} q^{\la_1+\la_2} \;, \eq$$
onto positive powers of the $a_i$'s, ensuring thereby the three inequalities: $\la_1\geq \la_2$, $\la_2\geq 1$ and $\la_1\geq \ell$.
It is convenient to  introduce the
MacMahon projection symbol 
$\Omega$, defined by
$$\Ol\ \,\sum_{n=-\y}^\y c_na^n = \left. \sum_{n\geq0} c_na^n\right|_{a=1}= \sum_{n\geq0} c_n\eq$$
and make use of 
 identities of the following type:
$$\ \Ol\ \,{1\over (1-a q) (1-a^{-1} q)} = \Ol\ \,{1\over (1-q^2)} \left({1\over
1-a q} + {a^{-1} q\over 1-a^{-1}q}\right) = {1\over (1-q)(1-q^2)} \;. \eq$$
The first two projections are rather direct  $$
\Olaaa\, \Olaa\, \Ola \, {a_3^{-\ell} a_2^{-1}\over (1-a_1a_3 q)(1-a_2 q/a_1)} = 
\Olaaa\, \Olaa\, {a_3^{-\ell} a_2^{-1}\over (1-a_3 q)(1-a_2a_3 q^2)}= 
\Olaaa\,{a_3^{1-\ell} q^2\over (1-a_3 q)(1-a_3 q^2)}\;, \eq$$
and for the third one, we have
$$
\eqalign{\Ol\,{a^{1-\ell} q^2\over (1-a q)(1-aq^2)}&= {a^{1-\ell} q^2\over (1-aq)(1-a q^2)}\left.\left(1-(1-aq)(1-aq^2)\sum_{p=0}^{\ell-2}\sum_{j=0}^p a^{p}q^{p+j}\right)\right|_{a=1}\cr& = {q^{\ell+1}+q^{\ell+2}-q^{2\ell+1}\over (1-q)(1-q^2)} \cr}\;. \eqlabel\gftwop$$
The $a_j$ are then fixed by comparing (\gftwo) and (\gftwop),  the solution of which being
$$a_j = {\rm max}\, (\ell-j,0)\;, \eq$$
as announced (cf. (\deCL)). This is valid for $0\leq \ell\leq k$.

\appendix{C}{The analysis of boundary terms via recurrence relations for generating functions}

We reconsider the construction of the generating functions for partitions $\la=(\la_1,\cdots,\la_N)$ into $N$ parts satisfying $\la_i\geq \la_{i+1}$ and $\la_i\geq \la_{i+2}+2r $, together with the boundary condition
$ \la_{N-1}\geq \ell $.
 The set of such partitions can be described schematically as (see e.g., [\FJMb])
$$\cdots(2r+\ell)(2r+1)(\ell)(1)^+\,,\eqlabel\pict$$
indicating that we build up these restricted partitions on the above ground state and  the $+$ sign indicates the position from  which we start the building up process (from right to left) by addition of ordinary partitions.  We then use this pictural representation to write down the recurrence relation between sets with different boundary conditions:
$$\eqalign{  \cdots  (2r+\ell)(2r+1)(\ell)(1)^+ -\cdots (2r+\ell+1)(2r+1)(\ell+1)(1)^+ &\cr   =  \quad  
\cdots (2r+\ell)(2r+1)^+(\ell)(1) + \cdots (2r+\ell)(2r+2)^+(\ell)(2) &\cr \qquad +\dots +
\cdots (2r+\ell)(2r+\ell)^+(\ell)(\ell)& \cr}\eq$$
The difference on the left-hand side generates the set of partitions for which the penultimate part is $\ell$ and this set is then broken, on the right-hand side,  into sets with prescribed values of the last two entries.
Denote by $p_\ell(w,N)$  the number of  partitions of weight $w$ (where $w=\sum \la_i$) with $N$ parts in the set (\pict). The above recurrence relation can be translated into the following condition
$$
p_{r, \ell}(w,N)-p_{r,\ell+1}(w,N)= \sum_{s=0}^{\ell-1}p_{r,\ell-s}(w-(2r+s)(N-2)-\ell-s-1,N-2)
\eq$$
On the left-hand side, we have used the observation that the set with the tail $(\ell) (s+1)$ (for $0\leq s\leq \ell-1$) is in one-to-one correspondence with the set obtained by deleting the last two parts $(\ell) (s+1)$ and subtracting $2r+s$ from each of the $N-2$ remaining parts, whose cardinality is thereby given  by $p_{r,\ell-s}(w-(2r+s)(N-2)-\ell-s-1,N-2)$. In terms of  the generating function (\gfr),
the above recurrence relation implies:
$$G_{\ell+1}(z,q) = G_\ell(z,q) - z^2 \sum_{s=0}^{\ell-1} q^{\ell+s+1}\, G_{\ell-s}(zq^{2r+s},q)\;. \eq$$
To this recurrence relation, we add the boundary condition $G_1=G_0$, with $G_0$ given in [\FJM].
In this way, we can construct $G_\ell$ recursively out of the known expression for $G_0$. In the following, we will only need the specialized version at $z=1$:
$$G_{r,\ell+1}(1,q) = G_{r,\ell}(1,q) -  \sum_{s=0}^{\ell-1} q^{\ell+s+1}\, G_{r,\ell-s}(q^{2r+s},q) \;. \eqlabel\recuG$$

Let us  now  show that for $0\leq \ell\leq k$, this relation leads to the expression already presented in (\genf),  (\deB) and (\deCL).  Take some $\ell+1\leq k$ and start by considering the difference
$$ G_{r,\ell}^{(0)}(1,q)\equiv G_{r,\ell}(1,q)-   q^{\ell+1}\, G_{r,\ell}(q^{2r},q)\eq $$
We first break the multisum expression for  $G_\ell(1,q)$ into two parts: one with $m_1=0$ and the other with  $m_1>0$. In the second sum, we redefine $m_1=m_1'+1$. The resulting expression has the same numerator as  $q^{\ell+1}\, G_{r,\ell}(q^{2r},q)$ so that their difference is easily computed.  Recombining the result with the  first multisum corresponding to $m_1=0$ yields an expression, denoted $G_{r,\ell}^{(0)}(1,q)$ above,  that is identical to $G_{r,\ell}(1,q)$ except that the linear coefficient of $m_1$ has been increased by 1. Next let us consider the difference between this expression $G_{r,\ell}^{(0)}(1,q)$ and the second term ($s=1$) of the sum of the right-hand side of (\recuG):
$$G_{r,\ell}^{(1)}(1,q)\equiv G_{r,\ell}^{(0)}(1,q))-   q^{\ell+2}\, G_{r,\ell}(q^{2r+1},q)\, .\eq$$
The same manipulations but with $m_1$ replaced by $m_2$ gives back $G_{r,\ell}^{(0)}(1,q)$ except that the linear coefficient of $m_2$ has been increased by 1. Proceeding in this way by successively taking into account the different terms of the sum, we end up with the prescription that going from $G_{r,\ell}$ to $G_{r,\ell+1}$, amount to increase by 1 all the linear coefficients of $m_1,\cdots, m_{\ell-1}$ and $m_\ell$ by 1 without  modifying the other mode coefficients. This demonstrates that the difference between the linear term without boundary condition, pertaining to $\ell=0,1$, and those with $\la_{N-1}\geq \ell$, with $1\leq \ell\leq k$,  is precisely given by the term max $(\ell-j)$ in (\deCL). 

By construction, this recombination of the right-hand side of (\recuG) works for $\ell+1\leq k$. This  bound can be  justified by the argument presented in appendix B concerning the independence of the coefficient of $m_k$ upon $\ell$: there are thus no more coefficient available to reshuffle.To see explicitly how this recombination process breaks down for $\ell>k$, it suffices to consider $\ell=3$ and $p=8$ (so that $k=2$). This suggests that for $\ell>k$, the generating function $G_{r,\ell}(z,q)$ might not be expressible in terms of a single  multiple sum.

\vskip0.3cm

\noindent {\bf ACKNOWLEDGMENTS}

The work of PJ is supported by EPSRC and partially  by the EC
network EUCLID (contract number HPRN-CT-2002-00325), while that of  PM is supported  by NSERC.

\vskip0.3cm

\centerline{\bf REFERENCES}
\immediate\closeout\refs \vskip 0.5cm
  \message{References}\input references

\end

The M(3,p) minimal models are reconsidered from the point of view of the extended algebra whose generators are the energy-momentum  tensor and  the primary field \phi_{2,1} of dimension $(p-2)/4$.
Within this framework, we provide a quasi-particle description of these models, in which all states are expressed solely in terms of the \phi_{2,1}-modes. More precisely, 
we show that all the states can be written in terms of \phi_{2,1}-type highest-weight states and their phi_{2,1}-descendants. We further demonstrate that the conformal dimension of these highest-weight states can be calculated from the \phi_{2,1}  commutation relations, the highest-weight conditions and associativity. For the simplest models (p=5,7), the full spectrum is explicitly reconstructed along these lines. For $p$ odd, the commutation relations between the \phi_{2,1} modes take the form of infinite sums, i.e., of  generalized commutation relations akin to parafermionic models. In that case, an unexpected  operator, generalizing the Witten index, is unravelled in the OPE of \phi_{2,1} with itself.
A quasi-particle basis formulated in terms of the sole \phi_{1,2} modes is  studied for all allowed values of p. We argue that it is governed by jagged-type partitions
further subject a difference 2 condition at distance 2. 
We demonstrate the correctness of this basis by constructing its generating function, from which the proper fermionic expression of the combination of the Virasoro irreducible characters \chi_{1,s} and \chi_{1,p-s} (for 1\leq s\leq [p/3]+1) are recovered.
As an aside, a practical technique for implementing associativity at the level of mode computations is presented, together with a general discussion of the relation between associativity and the Jacobi identities.